\documentclass[useAMS,usedcolumn,usegraphicx,usenatbib]{mn2e}
\usepackage{amssymb,amstext,amsfonts} %% ... with default font
\usepackage[fleqn]{amsmath}
\usepackage{subfigure}
\usepackage{slashed}
\usepackage[lowtilde]{url}
\usepackage{mathrsfs}
\usepackage{dcolumn}
\usepackage{hyperref}

\newcommand{\eqnref}[1]{(\ref{eq:#1})}
\newcommand{\figref}[1]{Fig.~\ref{fig:#1}}
\newcommand{\Figref}[1]{Figure~\ref{fig:#1}}
\newcommand{\tabref}[1]{Table~\ref{tab:#1}}
\newcommand{\secref}[1]{Sec.~\ref{sec:#1}}

\newcommand{\apref}[1]{Appendix~\ref{ap:#1}}

\newcommand{\units}[1]{\ensuremath{~\mathrm{#1}}}

\newcommand{\sub}[1]{\ensuremath{_\mathrm{#1}}}

\newcommand{\dd}{\ensuremath{\mathrm{d}}}

\newcommand{\intd}[4]{\ensuremath{\int_{#1}^{#2}{#3}\,\dd{#4}}}
\newcommand{\recip}[1]{\ensuremath{\frac{1}{#1}}}

\newcommand{\innerprod}[2]{\ensuremath{\left({#1}\middle|{#2}\right)}}

\title[EMRBs from extragalactic sources]{Extreme-mass-ratio-bursts from extragalactic sources}
\author[C.\ P.\ L.\ Berry and J.\ R.\ Gair]{C.\ P.\ L.\ Berry$^{1}$\thanks{E-mail: cplb2@cam.ac.uk} and J.\ R.\ Gair$^{1}$\\
$^{1}$Institute of Astronomy, University of Cambridge, Madingley Road, Cambridge, CB3 0HA, UK}

\begin{document}

\date{\today}

\pagerange{\pageref{firstpage}--\pageref{lastpage}} \pubyear{2013}

\maketitle

\label{firstpage}

\begin{abstract}
Extreme-mass-ratio bursts (EMRBs) are a class of potentially interesting gravitational wave signals. They are produced when a compact object passes through periapsis on a highly eccentric orbit about a much more massive object; we consider stellar mass objects orbiting the massive black holes (MBHs) found in galactic centres. Such a system may emit many EMRBs before eventually completing the inspiral. There are several nearby galaxies that could yield detectable bursts. For a space-borne interferometer like the \textit{Laser Interferometer Space Antenna}, sensitivity is greatest for EMRBs from MBHs of $\sim10^6$--$10^7 M_\odot$, which could be detected out to $\sim 100\units{Mpc}$. Considering the examples of M32, NGC 4945 and NGC 4395 we investigate if extragalactic EMRB signals can provide information about their sources. This is possible, but only if the periapse radius of the orbit is small, of the order of $r\sub{p} \lesssim 8 r\sub{g}$, where $r\sub{g} = GM c^{-2}$ is the gravitational radius of the MBH. This limits the utility of EMRBs as an astronomical tool. However, if we are lucky, we could place constraints on the mass and spin of nearby MBHs with $1\%$ precision.
\end{abstract}


\section{Introduction}\label{sec:Intro}

It is well established that space is big \citep[chapter 8]{Adams1979}. The Milky Way, our own island universe, is but one of a multitude of galaxies. Each one of these may have a massive black hole (MBH) nestled at its core \citep{Lynden-Bell1971, Soltan1982}.

In previous work \citep{Berry2013}, we considered measuring the properties of the Galaxy's MBH using extreme-mass-ratio bursts (EMRBs). An EMRB is a short gravitational wave (GW) signal produced when a small object passes through periapsis on an orbit about a much more massive body; in our case this is a stellar mass compact object (CO) orbiting the MBH. If the periapse radius of the orbit is sufficiently small ($r\sub{p} \lesssim 10 r\sub{g}$ for a $10 M_\odot$ CO, where $r\sub{g} = GMc^{-2}$ is a gravitational radius), a single burst can be highly informative about the MBH, improving our knowledge of its mass and spin.

EMRBs could be considered as the precursors to the better studied extreme-mass-ratio inspirals (EMRIs; \citealt{Amaro-Seoane2007}). Close encounters in the dense nuclear cluster surrounding the MBH scatter COs on to highly eccentric orbits. They proceed to emit an EMRB each orbit \citep*{Rubbo2006}. If they survive for long enough without being scattered again, the loss of the energy-momentum carried away by gravitational radiation leads the orbit to circularise; eventually the GW signal changes, so there is continuous significant emission and we have an EMRI. This continues until the inevitable plunge into the MBH. EMRBs are much shorter in duration than EMRIs; these may generate $\sim10^5$ cycles whereas bursts only generate $\sim1$. EMRBs do not have as much time to accumulate high signal-to-noise ratios (SNRs) and consequently are neither detectable to the same range, nor as informative as EMRIs. However, such an extreme-mass-ratio system could emit many bursts before transitioning to the EMRI regime, making EMRBs an interesting potential signal for GW detection.

In this work, we consider if EMRBs are detectable from other nearby galaxies. If so, they may be useful for constraining the properties of those galaxies' MBHs. Observations have shown that MBH masses are correlated with properties of the host galaxies, such as bulge luminosity, mass, velocity dispersion and light concentration \citep[e.g.,][]{Kormendy1995, Magorrian1998, Graham2011}. The two are linked via their shared history, such that one can inform us about the other.

Astrophysical black holes (BHs) are described by two quantities: mass $M$ and (dimensionless) spin $a_\ast$ \citep{Chandrasekhar1998}. The spin is related to the angular momentum $J$ by
\begin{equation}
a_\ast = \frac{cJ}{GM^2},
\end{equation}
and spans the range $0 \leq |a_\ast| < 1$. For many MBHs in the local neighbourhood, we have existing mass estimates. Measuring the spins would give us a complete picture, and would crucially give an insight into the formation histories of the galaxies \citep{Dotti2013,Volonteri2012a}.

MBHs accumulate mass and angular momentum through accretion and mergers \citep{Volonteri2010, Yu2002}; the spin encodes information about the mechanism that has most recently dominated the evolution. Accretion from a massive gaseous disc spins up the MBH, resulting in high spin values \citep{Volonteri2005}; randomly orientated accretion events lead to low spin values \citep*{King2006, King2008}; minor mergers with smaller BHs can decrease the spin \citep*{Hughes2003, Gammie2004} and major mergers between MBHs give a likely spin $|a_\ast| \sim 0.7$ \citep{Berti2008, Gonzalez2007}. Determining how the spin evolved shall provide clues about how the galaxy evolved \citep{Barausse2012}.

We have some MBH spin measurements from X-ray observations of active galactic nuclei \citep[e.g.,][]{Walton2013}. Estimates span the entire range of allowed values, but are typically in the range of $|a_\ast| \sim 0.7$ and above, with uncertainties of $\sim 10\%$. There may be an observational bias towards high spin values \citep{Brenneman2011}. It would be interesting to compare observations of the active galactic nuclei population with measurements from the population of nearby galaxies to see if they share universal characteristics or define distinct demographics. %Nardini2011, Patrick2011, Gallo2011, Lohfink2012

EMRBs could be an interesting signal for a space-borne GW detector, such as the \textit{Laser Interferometer Space Antenna} (\textit{LISA}; \citealt{Bender1998, Danzmann2003}) or the \textit{evolved Laser Interferometer Space Antenna} (\textit{eLISA}; \citealt{Jennrich2011, Amaro-Seoane2012a}).\footnote{The revised \textit{eLISA} concept is the same revised design as the \textit{New Gravitational-wave Observatory} (\textit{NGO}) submitted to the European Space Agency for their L1 mission selection.} At the time of writing, there is currently no funded mission. However, \textit{LISA Pathfinder}, a technology demonstration mission, is due for launch in 2015 \citep{Anza2005, Antonucci2012, McNamara2013}. There is optimism that a full mission shall follow in the subsequent decade. Since there does not exist a definite mission design, we stick to the classic \textit{LISA} design for the majority of this work, although we do use the \textit{eLISA} design when considering the detectability of bursts. The principal effect of using a descoped design is a reduction in SNR; this would reduce the precision to which parameters could be inferred.

EMRB waveforms are calculated and analysed as in \citet{Berry2013} and therefore, in the following, we give only an outline of the techniques used. Waveform construction and the numerical kludge approximation are outlined in \secref{Wave}. The basics of signal analysis are introduced in \secref{Sig}. In \secref{SNR}, the detectability of EMRBs from extragalactic MBHs is discussed. We show that bursts from other galaxies could be detected with \textit{LISA} or \textit{eLISA}. Following this, in \secref{Infer} we discuss how to assess information that could be extracted from these signals and in \secref{Res}, we present examples of the constraints we could place using EMRBs. We conclude in \secref{End} with a discussion of our findings.

We do not discuss in detail the question of event rates, which we defer to future work \citep{Berry2013b}. To be a useful tool for astronomy EMRBs must be both informative and sufficiently prevalent that they can be observed during a mission lifetime. We do not expect EMRBs to be common as they are only detectable across a small range of periapses. However, the larger the number of galaxies from which detectable EMRBs can originate, the higher the total event rate. If we find there are many galaxies hosting candidate sources, then we may be more confident that we could expect to detect at least one extragalactic burst. Our preliminary results suggest that we could expect $\sim 0.2$ extragalactic bursts per year per Milky Way equivalent galaxy. Therefore, while we expect that we cannot rely on EMRBs providing information about any particular source, we could learn something about a small subset of the candidates, those which do produce bursts during a mission. It seems unlikely that EMRBs could be as useful as EMRIs, but they could be a bonus source of information.

\section{Waveform generation}\label{sec:Wave}

To build waveforms we employ a semirelativistic approximation \citep{Ruffini1981}: the CO travels along a geodesic in Kerr spacetime, but radiates as if it were in flat spacetime. This approach is known as a numerical kludge (NK). Comparison with more accurate, and computationally intensive, methods has shown that NK waveforms are reasonably accurate for extreme-mass-ratio systems (\citealt*{Gair2005}; \citealt{Babak2007}): typical errors are a few percent \citep{Tanaka1993,Gair2005,Berry2013}. Binding the motion to a true geodesic ensures that the signal has the correct frequency components, although neglecting the effects of background curvature means that these do not have the correct amplitudes. The geodesic parameters are kept fixed throughout the orbit, as there should be negligible evolution due to the emission of gravitational radiation.

All bursts are assumed to come from marginally bound, or parabolic, orbits. In this case, the CO starts at rest at infinity and has a single passage through periapsis. If the periapse radius is small enough, the orbit may still complete a number of rotations about the MBH; these are zoom--whirl orbits \citep{Glampedakis2002a} which zoom in from large radius, complete several rapid rotations about the MBH and then zoom out again.

When integrating the Kerr geodesic equations, we use angular variables instead of the radial and polar Boyer--Lindquist coordinates \citep{Drasco2004}
\begin{align}
r = {} & \frac{2 r\sub{p}}{1 + \cos\psi};\\
\cos^2\theta = {} & \frac{Q}{Q+L_z^2}\cos^2\chi = \sin^2 \iota \cos^2\chi,
\end{align}
where $r\sub{p}$ is the periapse radius, $Q$ is the Carter constant, $L_z$ is the angular momentum about the $z$-axis and $\iota$ is the orbital inclination \citep*{Glampedakis2002}. This parametrization avoids complications associated with turning points of the motion.

Once the geodesic is constructed, we identify the Boyer--Lindquist coordinates with flat-space spherical polar coordinates \citep{Gair2005, Babak2007}. This choice is not unique, as a consequence of the arbitrary nature of the NK approximation. Using flat-space oblate spheroidal coordinates gives quantitatively similar results \citep{Berry2013}. The quadrupole--octupole formula is used to derive the gravitational strain \citep{Bekenstein1973, Press1977, Babak2007, Yunes2008}. The inclusion of higher order terms modifies the amplitudes of some frequency components for the more relativistic orbits by a few tens of percent although the overall integrated effect is smaller.

The waveform is specified by a set of $14$ parameters, which are as follows.
\begin{enumerate}
\item[(1)] The MBH's mass $M$.
\item[(2)] The spin parameter $a_\ast$.
\item[(3, 4)] The orientation angles for the MBH spin $\Theta\sub{K}$ and $\Phi\sub{K}$.
\item[(5)] The source distance $R$ divided by the CO mass $\mu$, which we denote as $\zeta = R/\mu$. This scales the amplitude of the waveform.
\item[(6, 7)] The angular momentum of the CO parametrized in terms of total angular momentum $L_\infty = \sqrt{Q + L_z^2}$ and inclination $\iota$.
\item[(8--10)] The angular phases at periapse, $\phi\sub{p}$  and $\chi\sub{p}$ (which determines $\theta\sub{p}$), and the time of periapse $t\sub{p}$.
\item[(11, 12)] The coordinates of the source. Sky position is already determined to high accuracy for each galaxy. Since an EMRB can only give weak constraints on source position we take it as known and do not infer it.
\item[(13, 14)] The orbital position of the detector. This should be known and need not be inferred. We assume the same initial position as \citet{Cutler1998}; this does not qualitatively influence results.
\end{enumerate}
We are interested in inferring the first $10$. The most interesting are the MBH's mass and spin.

We have assumed that sky position is known; to be able to do this in practice we must be able to successfully identify the source galaxy. No work has yet been done on EMRB detection and we will defer developing a detection algorithm for future work. However, we shall see that there are only a few potential galaxies that could produce detectable EMRBs. It should therefore not be too computationally expensive to check all the candidate sky positions. If multiple galaxies lie close together on the sky, such that they cannot be distinguished, it could be possible to use constraints on the MBH mass to differentiate them. This would not help with galaxies for which we do not have good MBH mass estimates.

\section{Signal analysis}\label{sec:Sig}

In this section we briefly cover the basics of GW analysis. A more complete discussion can be found in \citet{Finn1992} and \citet{Cutler1994}, and a review in \citet{Jaranowski2012}; those familiar with the subject may skip this section. In the following, detectors are labelled with indices $A = \{\mathrm{I}, \mathrm{II}\}$ for \textit{LISA}, which has three arms and acts as two detectors, and $A = \{\mathrm{I}\}$ for \textit{eLISA}, which has only two arms and so acts as a single detector.

The measured strain $\boldsymbol{s}(t)$ is the combination of the signal and the detector noise
\begin{equation}
\boldsymbol{s}(t) = \boldsymbol{h}(t) + \boldsymbol{n}(t);
\end{equation}
we assume the noise is stationary and Gaussian, and noise in multiple data channels is uncorrelated, but shares the same spectral characterisation \citep{Cutler1998}. We can then define a signal inner product \citep{Cutler1994}
\begin{equation}
\innerprod{\boldsymbol{g}}{\boldsymbol{k}} = 2\intd{0}{\infty}{\frac{\tilde{g}_A^\ast(f)\tilde{k}_A(f) + \tilde{g}_A(f)\tilde{k}_A^\ast(f)}{S\sub{n}(f)}}{f},
\label{eq:inner}
\end{equation}
introducing the Fourier transform
\begin{equation}
\tilde{g}(f) = \mathscr{F}\{g(t)\} = \intd{-\infty}{\infty}{g(t)\exp(2\pi i ft)}{t}
\end{equation}
and noise spectral density $S\sub{n}(f)$. We use the noise model of \citet{Barack2004} for \textit{LISA} and the simplified sensitivity model from \citet{Jennrich2011} for \textit{eLISA}.

The SNR is
\begin{equation}
\rho[\boldsymbol{h}] = \innerprod{\boldsymbol{h}}{\boldsymbol{h}}^{1/2}.
\label{eq:SNR}
\end{equation}
The probability of a realization of noise $\boldsymbol{n}(t) = \boldsymbol{n}_0(t)$ is
\begin{equation}
p(\boldsymbol{n}(t) = \boldsymbol{n}_0(t)) \propto \exp\left[-\recip{2}\innerprod{\boldsymbol{n}_0}{\boldsymbol{n}_0}\right].
\end{equation}
Therefore, if the incident waveform is $\boldsymbol{h}(t)$, the probability of measuring signal $\boldsymbol{s}(t)$ is
\begin{equation}
p(\boldsymbol{s}(t)|\boldsymbol{h}(t)) \propto \exp\left[-\recip{2}\innerprod{\boldsymbol{s}-\boldsymbol{h}}{\boldsymbol{s}-\boldsymbol{h}}\right].
\label{eq:sig_prob}
\end{equation}

\section{Detectability}\label{sec:SNR}

The detectability of a burst is determined by its SNR. We assume a detection threshold of $\rho = 10$. The SNR of an EMRB depends upon many parameters; for a given MBH, the most important is the periapse radius $r\sub{p}$. There is a good correlation between $\rho$ and $r\sub{p}$; other parameters specifying the inclination of the orbit; the orientation of the system with respect to the detector, or the MBH spin only produce scatter about this trend. The form of the $\rho$--$r\sub{p}$ relation depends upon the noise curve.

We parametrize the detectability in terms of a characteristic frequency $f_\ast$. The speed at periapse scales like $v \sim \sqrt{GM/r\sub{p}}$; the characteristic time taken for the position to change is then $T \sim r\sub{p}/v$, and so we define the characteristic frequency as
\begin{equation}
f_\ast = \sqrt{\frac{GM}{r\sub{p}^3}}.
\end{equation}
This allows comparison between different systems where the same periapse does not correspond to the same frequency and thus the same point of the noise curve.

We also expect the SNR to scale with other quantities. We define a characteristic strain amplitude for a burst $h_0$; we expect $\rho \propto h_0$, where the proportionality will be set by a frequency-dependent function that includes the effect of the noise curve. Assuming that the strain is dominated by the quadrupole contribution (\citealt*[section 36.10]{Misner1973}; \citealt*[section 17.9]{Hobson2006}) we expect
\begin{equation}
h_0 \sim \frac{G}{c^6}\frac{\mu}{R}\frac{\dd^2}{\dd t^2}\left(r^2\right),
\end{equation}
where $\mu$ is the CO mass, $R$ is the distance to the MBH, $t$ is time and $r$ is a proxy for the position of the orbiting object. The characteristic rate of change is set by $f_\ast$ and the characteristic length-scale is set by $r\sub{p}$. Hence
\begin{align}
h_0 \sim {} & \frac{G}{c^6}\frac{\mu}{R}f_\ast^2 r\sub{p}^2 \\
 \sim {} & \frac{G^{5/2}}{c^6}\frac{\mu}{R}f_\ast^{-2/3}M^{2/3}.
\end{align}
Using this, we can factor out the most important dependences to give a scaled SNR defined by
\begin{equation}
\rho_\ast = \left(\frac{\mu}{M_\odot}\right)^{-1}\left(\frac{R}{\mathrm{Mpc}}\right)\left(\frac{M}{10^6 M_\odot}\right)^{-2/3}\rho.
\label{eq:SNR-scaling}
\end{equation}

Space-based detectors are most sensitive to extreme-mass-ratio signals originating from systems containing MBHs with masses $\sim10^6 M_\odot$. Higher mass objects produce signals at too low frequencies. We considered several nearby MBHs that were likely candidates for detectable burst signals. Details are given in \tabref{MBHs}.
\begin{table*}
%\begin{minipage}{\columnwidth}
 \centering
  \caption{Sample of nearby MBHs that are candidates for producing detectable EMRBs.\label{tab:MBHs}}
  \begin{tabular}{@{} l c  D{.}{.}{2.5} l @{}}
  \hline
   Galaxy & \multicolumn{1}{c}{$M/10^6 M_\odot$} & \multicolumn{1}{c}{$R/\mathrm{Mpc}$} & References \\
 \hline
 Milky Way (MW) & $4.31 \pm 0.36$ & 0.00833 & \citet{Gillessen2009} \\
 Andromeda (M31, NGC 224) & $140^{+90}_{-30}$ & 0.770 & \citet{Bender2005,Karachentsev2004} \\
 M32 (NGC 221) & $2.5 \pm 0.5$ & 0.770 & \citet{Verolme2002,Karachentsev2004} \\
 Circinus & $1.1 \pm 0.2$ & 2.82 & \citet{Graham2008,Greenhill2003,Karachentsev2007} \\
 NGC 4945 & $1.4^{+0.7}_{-0.5}$ & 3.82 & \citet{Greenhill1997,Ferrarese2005,Karachentsev2007} \\
 Sculptor (NGC 253) & $10^{+10}_{-5}$ & 3.5 & \citet{Graham2011,Rodriguez-Rico2006,Rekola2005} \\
 NGC 4395 & $0.36 \pm 0.11$ & 4.0 & \citet{Peterson2005,Thim2004} \\
 M96 (NGC 3368) & $7.3 \pm 1.5$ & 10.1 & \citet{Graham2011,Nowak2010,Tonry2001} \\
 NGC 3489 & $5.8 \pm 0.8$ & 11.7 & \citet{Graham2011,Nowak2010,Tonry2001} \\
\hline
\end{tabular}
%\end{minipage}
\end{table*}
For each, we calculated SNRs at $\sim 10^4$ different periapse distances, uniformly distributed in log space between the innermost orbit and $100 r\sub{g}$. Each had a spin and orbital inclination randomly chosen from distributions uniform in $a_\ast$ and $\cos \iota$.\footnote{The innermost orbit depends upon $a_\ast$ and $\iota$, hence these are drawn first.} For every periapse, five SNRs were calculated, each having a different set of ancillary parameters specifying the relative orientation of the MBH, the orbital phase and the position of the detector, drawn from appropriate uniform distributions.

\begin{figure}
\begin{center}
 \includegraphics[width=0.43\textwidth]{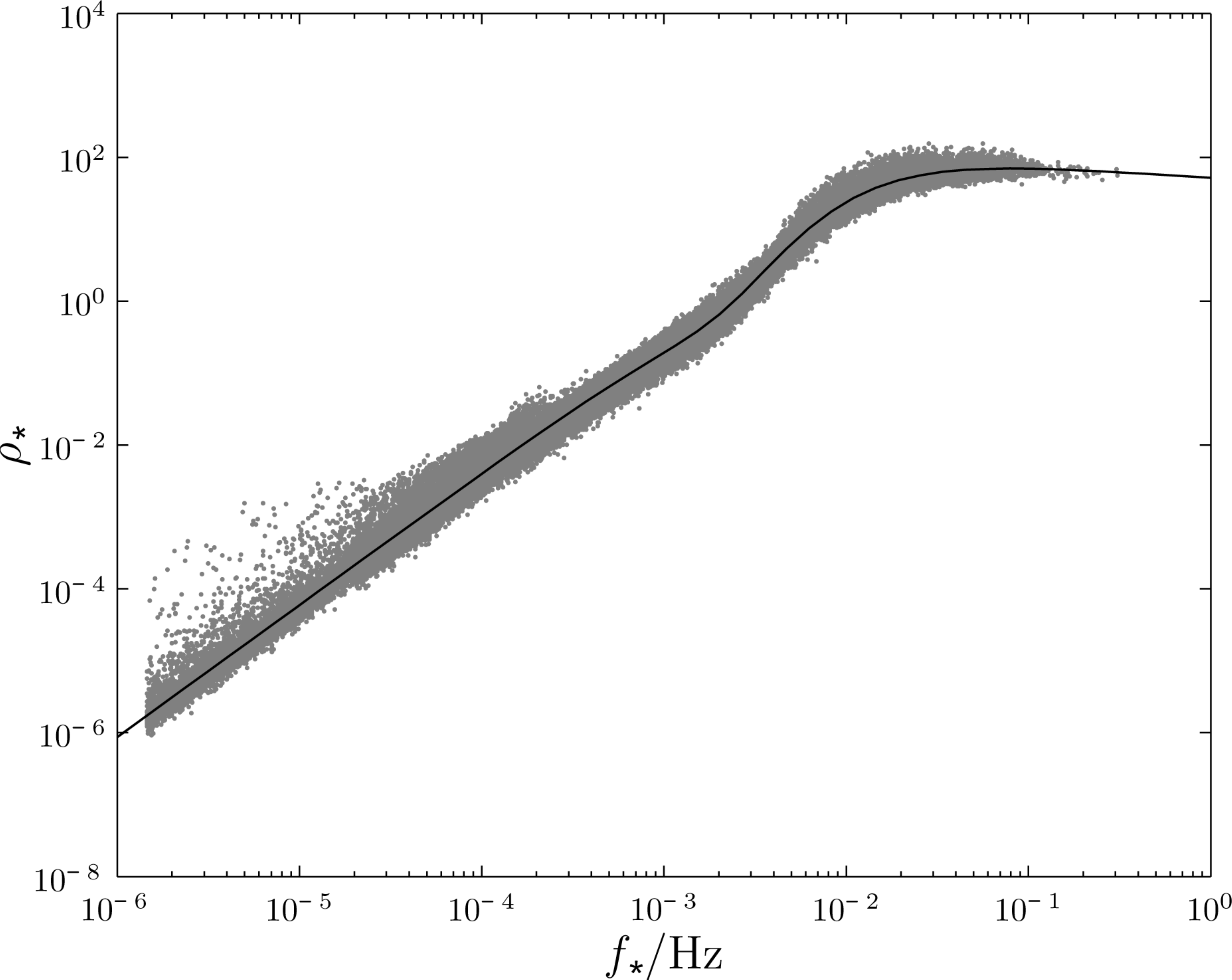}
 \caption{Scaled SNR for EMRBs as a function of characteristic frequency. The fitted curve from \eqnref{scaled-SNR} is indicated by the line.\label{fig:scaled-SNR}}
   \end{center}
\end{figure}

The scaled SNRs are plotted in \figref{scaled-SNR}. The plotted points are the average values of $\ln \rho_\ast$ calculated for each periapse distance.
The curve shows that EMRB SNR does scale as expected, and $\rho_\ast$ can be described as a one-parameter curve. There remains some scatter about this: the larger scatter at low frequencies is a consequence of numerical noise from dealing with very low SNRs from Andromeda; removing the averaging over ancillary parameters increases the scatter to be typically about an order of magnitude in total. However, the fit is good enough for rough calculations.

We approximate the trend with a parametrized curve
\begin{equation}
\rho_\ast = \alpha_1 \left(\frac{f_\ast}{\mathrm{Hz}}\right)^{\beta_1} \left[1 + \left(\alpha_2 \frac{f_\ast}{\mathrm{Hz}}\right)^{\beta_2}\right]\left[1 + \left(\alpha_3 \frac{f_\ast}{\mathrm{Hz}}\right)^{\beta_3}\right]^{-\beta_4}.
\label{eq:scaled-SNR}
\end{equation}
To fit this, we treat the problem as if it were a likelihood maximisation, with each averaged point having a Gaussian likelihood with standard deviation defined from the scatter because of the variation in the ancillary parameters. The optimised values for \textit{LISA} are
\begin{equation}
\begin{split}
&\alpha_1 \simeq 8.93 \times 10^4; \ \ \alpha_2 \simeq 4.68 \times 10^2; \ \  \alpha_3 \simeq 1.84 \times 10^2;\\
&\beta_1 \simeq 1.84; \ \ \beta_2 \simeq 3.23; \ \ \beta_3 \simeq 1.27; \ \ \beta_4 \simeq 4.13.
\end{split}
\end{equation}

Using our fitted trends, it is possible to invert \eqnref{SNR-scaling} to find the furthest distance that a system contain an MBH of a given mass can produce detectable bursts. In calculating the maximum SNR it is necessary to decide upon a maximum $f_\ast$. This corresponds to the minimum periapse radius which is in turn determined by the MBH spin. For the optimal case with a maximally rotating MBH, the innermost periapsis is $r\sub{p} = r\sub{g}$; for a non-rotating MBH, the innermost periapsis would be $r\sub{p} = 4r\sub{g}$. We shall use both as limits for the maximum SNR.

\begin{figure}
\begin{center}
 \includegraphics[width=0.43\textwidth]{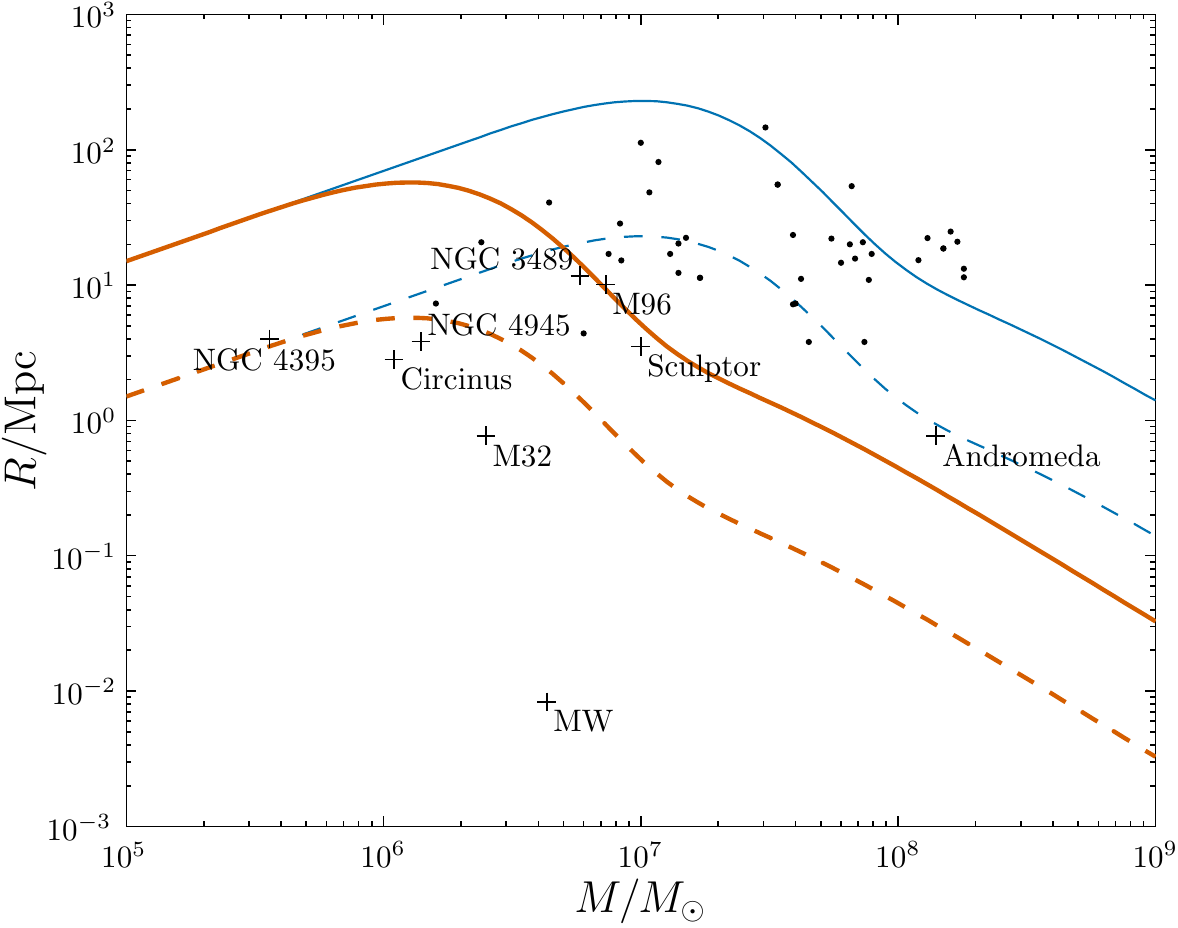}
 \caption{Limit of detection using \textit{LISA} for EMRBs originating from MBHs of mass $M$ and distance $R$ with CO of mass $\mu = 1 M_\odot$ (dashed line) or $\mu = 10 M_\odot$ (solid line). The detection threshold is assumed to be $\rho = 10$. The thicker line is the limit for non-rotating MBHs, the thinner is for maximally rotating MBHs. Sources below the relevant line are potentially detectable. The crosses indicate the selected sample of MBHs used to calibrate the curve and the dots indicate other nearby MBHs with known masses. The trends should not be extrapolated to lower MBH masses.\label{fig:detect}}
   \end{center}
\end{figure}

\Figref{detect} shows the detectability limit for $\mu = 1 M_\odot$ and $\mu = 10 M_\odot$ COs. In addition to the sample of MBHs from \tabref{MBHs} we plot additional nearby MBHs \citep[see][and references therein]{Graham2008,Graham2011,Graham2013}.
The more massive COs are detectable to a greater distance, but are also the more likely sources since mass segregation ensures that they are more likely to be on orbits that pass close to the MBH \citep{Bahcall1977, Alexander2009, Preto2010}. Limits using periapsis of $r\sub{g}$ and $4r\sub{g}$ are shown: intermediate spin values would have limits between these two. In any case, these are strict bounds; it is unlikely that we would observe a burst from the optimal orbit. Therefore bursts from MBHs outside the curve are impossible to detect and those inside may be possible, but need not be probable, to detect.

It appears that there are many extragalactic MBHs which could produce observable bursts. From the sample in \tabref{MBHs} all could be detected. Andromeda could only be detected if it has a high spin value. It is therefore less promising than the others. NGC 3489, M96 and Sculptor lie on the boundary of detectability for non-spinning sources with a $10 M_\odot$ CO. They are therefore of marginal interest: we do not necessarily need any special requirement for the spin, but such close orbits would be infrequent. NGC 4395, NGC 4945 and Circinus are around the boundary of detectability for a $1 M_\odot$ CO. Hence, we could potentially see bursts from white dwarfs or neutron stars as well as BHs. M32 is the best extragalactic source, lying safely within the detection limit for $1 M_\odot$ COs. Outside of our sample there are other MBHs with measured masses that are of interest. A great many could potentially be detected using optimal bursts from $10 M_\odot$ COs orbiting a maximally rotating MBH.\footnote{Many galaxies of the Virgo cluster fall in this category. This could potentially make identifying the source galaxy more difficult as the candidates are close together. Since we would have to be fortunate to encounter this problem, we shall not be overly concerned by it.} The most promising MBHs not included in our test sample are found in M64 (NGC 4826), NGC 3076 and M94 (NGC 4736).

\begin{figure}
\begin{center}
 \includegraphics[width=0.43\textwidth]{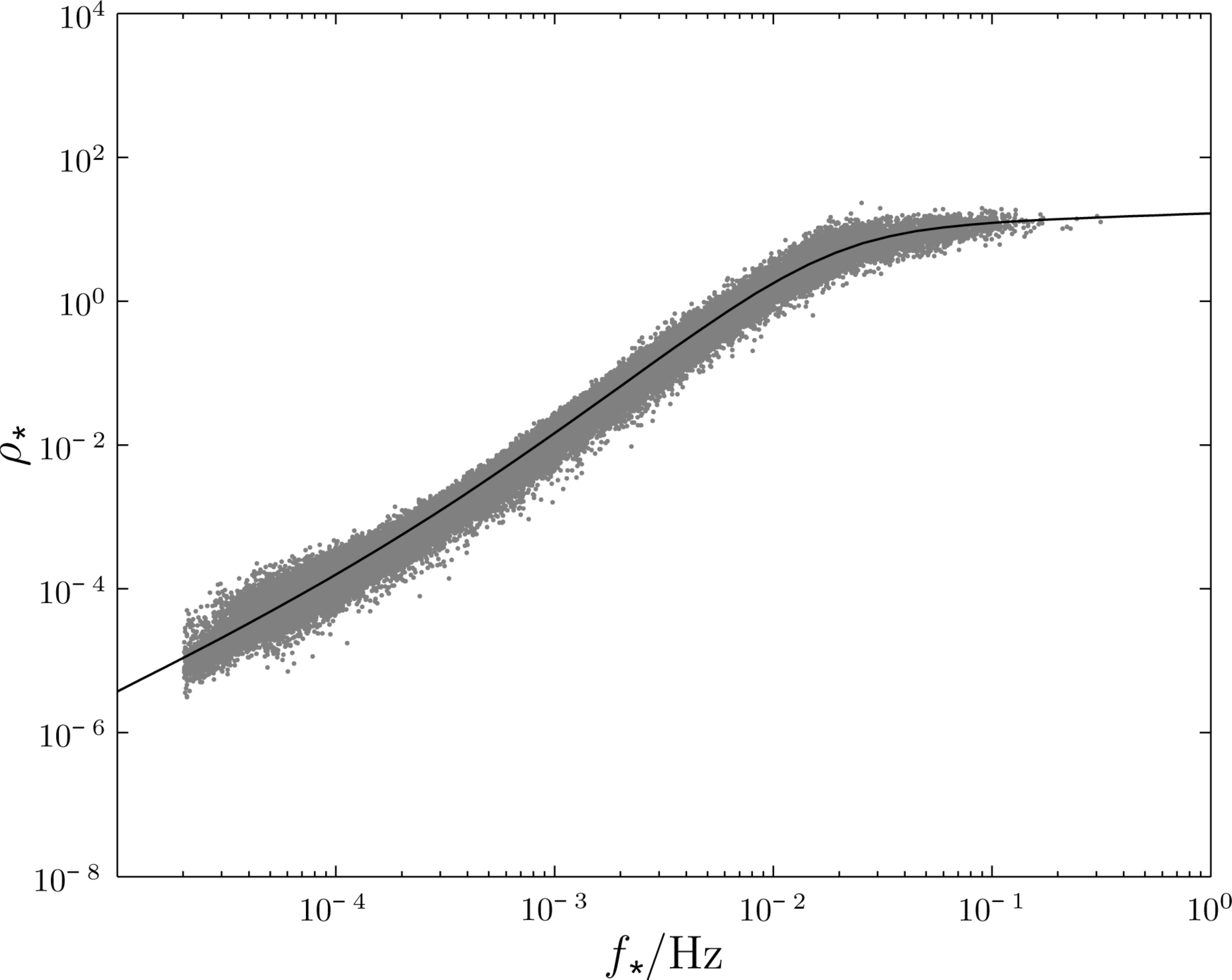}
 \caption{Scaled SNR for EMRBs as a function of characteristic frequency for the \textit{eLISA} design. The fitted curve from \eqnref{scaled-SNR} is indicated by the line.\label{fig:scaled-SNR-eLISA}}
   \end{center}
\end{figure}

We can repeat the analysis for \textit{eLISA}. The scaled SNRs are shown in \figref{scaled-SNR-eLISA}.
Since Andromeda was only marginally of interest for the classic \textit{LISA} design, we did not include it this time. This reduces the scatter at low characteristic frequencies.

The curve is fitted with
\begin{equation}
\begin{split}
&\alpha_1 \simeq 73.9; \ \ \alpha_2 \simeq 4.99 \times 10^3; \ \ \alpha_3 \simeq 52.7;\\
&\beta_1 \simeq 1.47; \ \ \beta_2 \simeq 0.85; \ \ \beta_3 \simeq 1.76; \ \ \beta_4 \simeq 1.25.
\end{split}
\end{equation}
The fit parameters are markedly different from those for \textit{LISA}. However, since we are fitting a phenomenological model and the parameters have no physical significance, we are not concerned with this. The parameters yield a good fit to the data, which is all that we are concerned about here.

\begin{figure}
\begin{center}
 \includegraphics[width=0.43\textwidth]{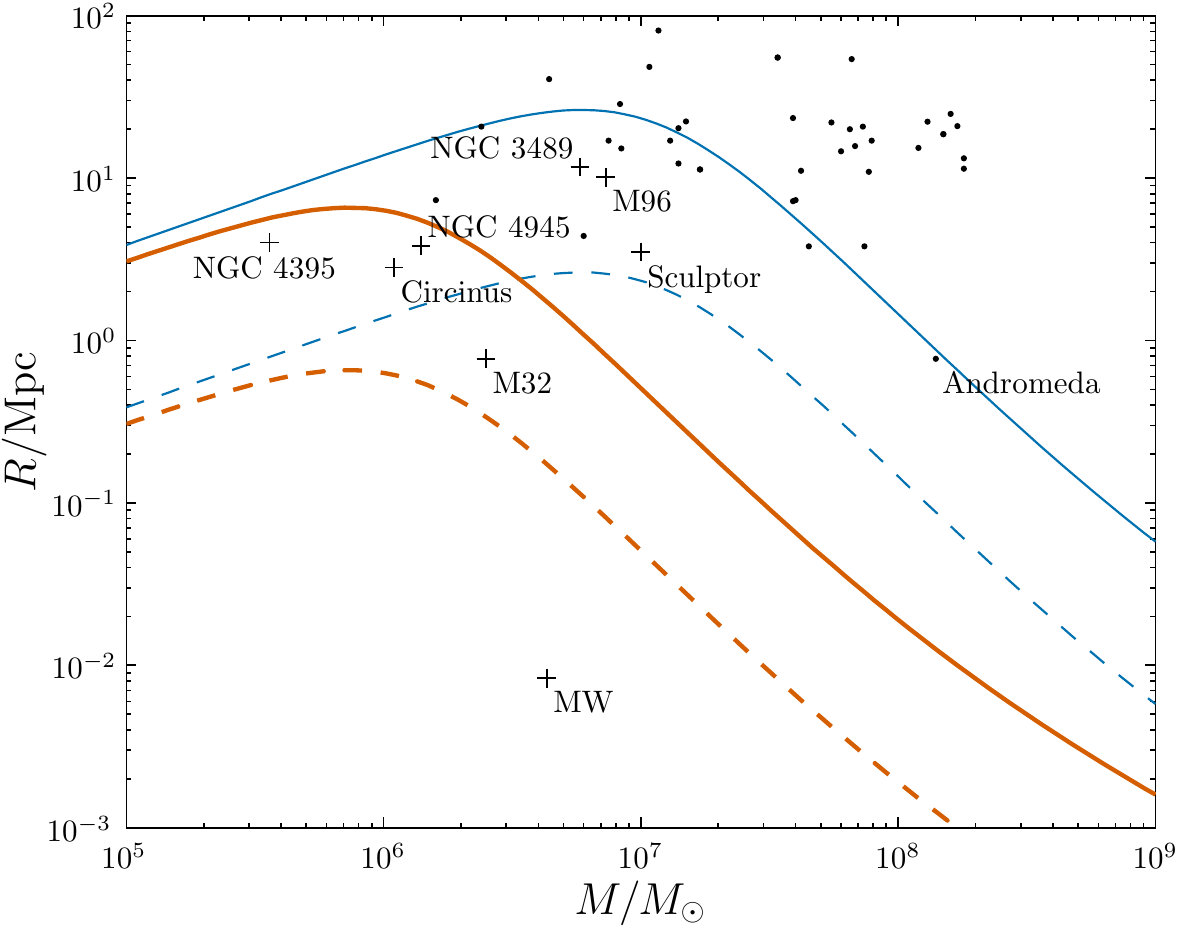}
 \caption{Limit of detection using \textit{eLISA} for EMRBs originating from MBHs of mass $M$ and distance $R$ with CO of mass $\mu = 1 M_\odot$ (dashed line) or $\mu = 10 M_\odot$ (solid line). The detection threshold is assumed to be $\rho = 10$. The thicker line is the limit for non-rotating MBHs, the thinner is for maximally rotating MBHs. Sources below the relevant line are potentially detectable. The crosses indicate the selected sample of MBHs used to calibrate the curve and the dots indicate other nearby MBHs with known masses. The trends should not be extrapolated to lower MBH masses.\label{fig:detect-eLISA}}
   \end{center}
\end{figure}

Using this fit to find the detectability range results in the curves shown in \figref{detect-eLISA}.
The maximum distances are reduced compared to the \textit{LISA} case indicating that detectable bursts would be much rarer. There still remain a number of potential candidate galaxies. From our sample, Andromeda is on the very edge of possibility. NGC 3489, M96 and Sculptor require a high spin, making them unlikely sources. NGC 4395, NGC 4945 and Circinus can be detected without the high spin assuming a $10 M_\odot$ CO. Of the extragalactic sources, only M32 remains detectable with a $1 M_\odot$ CO, and still it requires a non-zero spin.

Using either noise curve we see that EMRBs could potentially be seen from a range of galaxies. The Galaxy's MBH remains securely detectable in either case. M32 is the next best. MBHs with masses $\sim 10^6$--$10^7 M_\odot$ are observable to the greatest distance. We currently know of few MBHs with masses at the lower end of the spectrum ($10^5$--$10^6 M_\odot$) but these would be good potential candidates.

\section{Parameter inference}\label{sec:Infer}

We are not only interested in discovering if EMRBs are detectable, but also if we can extract information from the signals about their sources. The probability that the burst is described by parameters $\boldsymbol{\lambda}$ is given by the posterior distribution as determined from Bayes' theorem:
\begin{equation}
p(\boldsymbol{\lambda}|\boldsymbol{s}(t)) = \frac{p(\boldsymbol{s}(t)|\boldsymbol{\lambda})p(\boldsymbol{\lambda})}{p(\boldsymbol{s}(t))},
\end{equation}
where $p(\boldsymbol{s}(t)|\boldsymbol{\lambda})$ is the likelihood of the parameters, $p(\boldsymbol{\lambda})$ is the prior for the parameters, and $p(\boldsymbol{s}(t))$ is the evidence which is just a normalising factor for our purposes.

\subsection{Mapping the posterior}\label{sec:MCMC}

To discover if any parameters can be accurately inferred, we must characterise the shape of the posterior. Markov chain Monte Carlo (MCMC) methods are commonly used for inference problems \citep[chapter 29]{MacKay2003}. The parameter space is explored by the construction of a chain of random samples. A new point is accepted with a rate dependent upon its probability, such that the converged chain reflects the underlying distribution \citep{Metropolis1953,Hastings1970}. We employ the same semi-adaptive algorithm that was previously used in \citet{Berry2013}.\footnote{The only modification was to lower the target acceptance rate to $\sim0.08$. This appeared to give improved convergence for these distributions.} This follows the suggestion of \citet*{Haario1999} having an initial period where the proposal distribution (used in the selection of new points) is adjusted to match the distribution of points previously accepted, before proceeding to the main phases where the proposal is kept fixed in order to be purely Markovian.

The likelihood of a set of parameters is found using \eqnref{sig_prob}:
\begin{equation}
p(\boldsymbol{s}(t)|\boldsymbol{\lambda}) \propto \exp\left[-\recip{2}\innerprod{\boldsymbol{s}-\boldsymbol{h}(\boldsymbol{\lambda})}{\boldsymbol{s}-\boldsymbol{h}(\boldsymbol{\lambda})}\right].
\label{eq:likelihood}
\end{equation}
We assume non-informative priors on the parameters to reflect a state of ignorance: we do not incorporate information we have from other measurements so that the results indicate what could be learnt from an EMRB observation alone.

\subsection{Characterising uncertainty}

Having recovered the posterior distribution it is necessary to quantify the accuracy to which parameters could be measured. If the posterior were Gaussian, this can be done just by using the standard deviation. An alternative is to use the range that encloses a given probability, but this is misleading if the distribution is multimodal. A robust means of characterising the width is by using a $k$-dimensional ($k$-d) tree.

A $k$-d tree is a type of binary space partitioning tree \citep[sections 5.2, 12.1, 12.3]{Berg2008}. It is constructed by splitting the parameter space into two by finding the median point in one dimension. The two pieces are then split by finding their medians in another dimension. This continues recursively until the desired number of partitions, known as leaves, has been created. When applied to a sampled probability distribution, a $k$-d tree has smaller leaves in the regions of high probability which are of most interest \citep{Weinberg2012}. It builds a natural decomposition of the parameter space, giving a means of binning samples.

For a given probability $p$, the corresponding confidence region is the smallest area of parameter space in which we expect that the true values lie with that probability. A simple means of constructing a confidence range is to find the smallest combination of $k$-d tree leaves that contain the desired probability. To do this we rank the leaves by size; the smallest corresponds to the highest probability area and is the starting point for the confidence range. We continue adding the next smallest leaf until the total probability enclosed is $p$. Summing the areas of the leaves gives an estimate for the range.

However, this approach is biased. Whenever a random fluctuation in the sampling gives an excess of points in one area the overdensity leads to a smaller leaf size and then the preferential inclusion of that leaf in the confidence interval. Conversely, an underdensity leads to a larger leaf that is liable to be external to the confidence range. If there are a small number of points per leaf we shall overstate our confidence as the constructed range is too small.\footnote{This can be visualised by considering the simple example of dividing in two samples from a one dimensional uniform distribution. We would expect one partition to be more densely populated than the other because of random fluctuations, and we shall always pick this smaller leaf as our $p = 0.5$ confidence range. As the number of points increases we expect that this bias would decrease.}

Biasing may be avoided by using a two-step method which separates the creation and ordering of the partitions from the building of the confidence range \citep*{Sidery2013}. This is done by dividing our data into two disjoint random samples.\footnote{We split our data into two equal parts. This may not be the optimal rationing, but is a sensible first guess. Some preliminary experimentation shows that it is not too important, provided that the splitting is not too unbalanced. The point at which this occurs depends upon the underlying distribution.} The first is used to construct the $k$-d tree in the standard way. The leaves are then ordered by size. We then use the second set to populate the leaves. We again start with the smallest leaf and work down the ranking until the encompassed probability is $p$. The total range is the estimate for the $p$ confidence level.

The first step creates bins that are of appropriate resolution. We therefore have the benefit of using a $k$-d tree. By using an independent set of points to build the confidence level, we eliminate any bias because there should be no correlation in fluctuations between the two sets. Any leaves that are too small are expected to receive a below average number of points in the second step and any that are too large are expected to receive more. This corrects the expectation for the confidence level.

In this case, we are interested in the confidence levels for the marginalised distributions for each parameter. We therefore construct $1$-d trees, which are easily implemented. We have a large number of points and low dimensionality so biasing should not be an issue. To characterise our distributions we find the $p = 0.68$ confidence range and take the half-width of this.

\section{Results}\label{sec:Res}

To investigate the potential of extragalactic EMRBs, we considered a sample of bursts from M32, the most promising candidate; NGC N4945, which is near to the optimal mass for \textit{LISA} (without assuming spin), and NGC 4395, the lightest MBH in our sample. Circinus is similar to NGC 4945, so we expect comparable results: EMRBs from Circinus should be slightly more useful as Circinus is closer.

\subsection{Posterior forms}

As for bursts from the Galactic Centre, posteriors can show strong and complicated parameter degeneracies. The lower SNR compared to Galactic bursts yields wider distributions. As the periapsis increases, the posteriors deteriorate, becoming uninformative. Some example results are shown in \apref{posterior}.

The general trend is for bursts from orbits with smaller periapses to be narrower and more Gaussian. As the periapse increases, and SNR decreases, the distributions broaden becoming more non-Gaussian. Curving degeneracies and secondary modes develop. Eventually, the distribution broadens to encompass the entire permitted range for the spin and various angular parameters, effectively making these quantities unconstrained.

\subsection{Parameter uncertainties}

\begin{figure*}
\begin{center}
\subfigure[Natural logarithm of MBH mass.]{\includegraphics[width=0.43\textwidth]{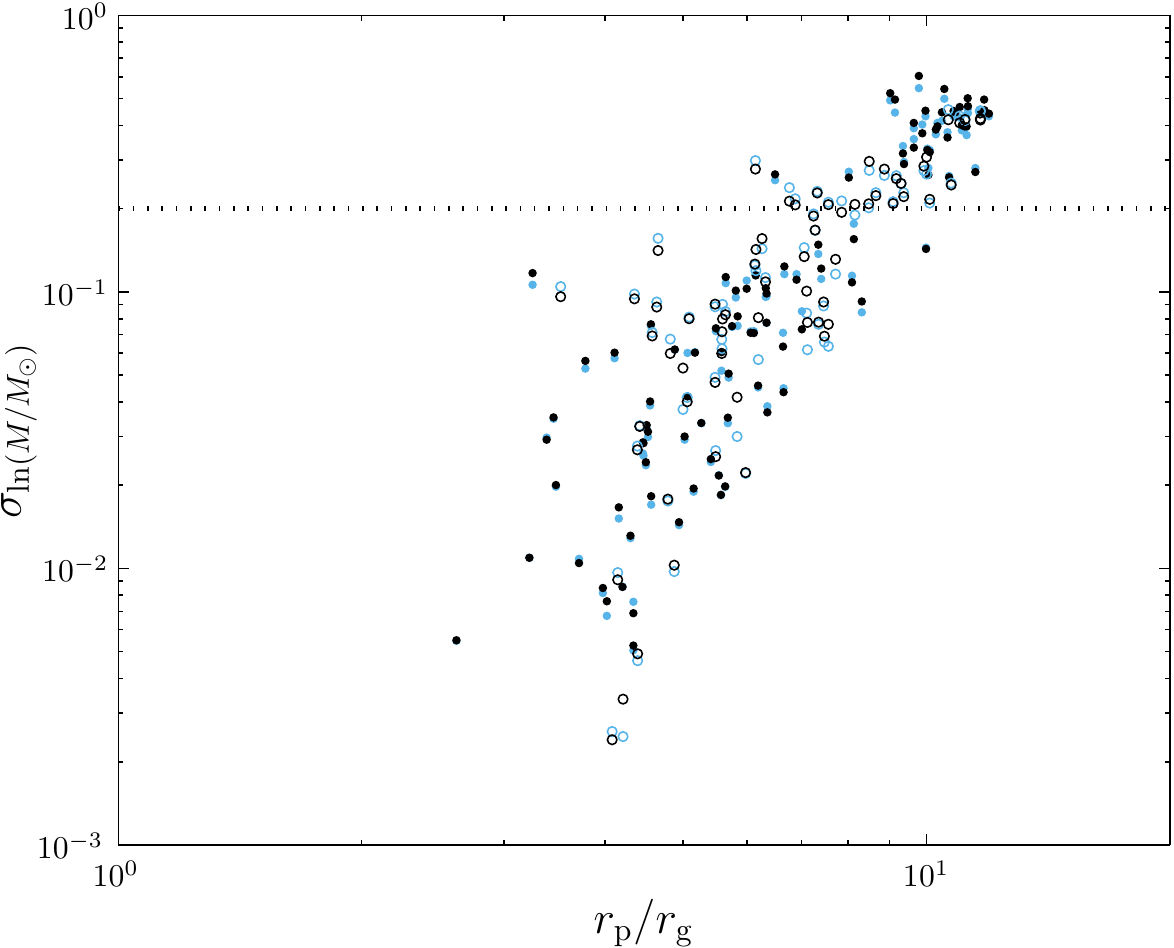}} \quad
\subfigure[Dimensionless spin.]{\includegraphics[width=0.43\textwidth]{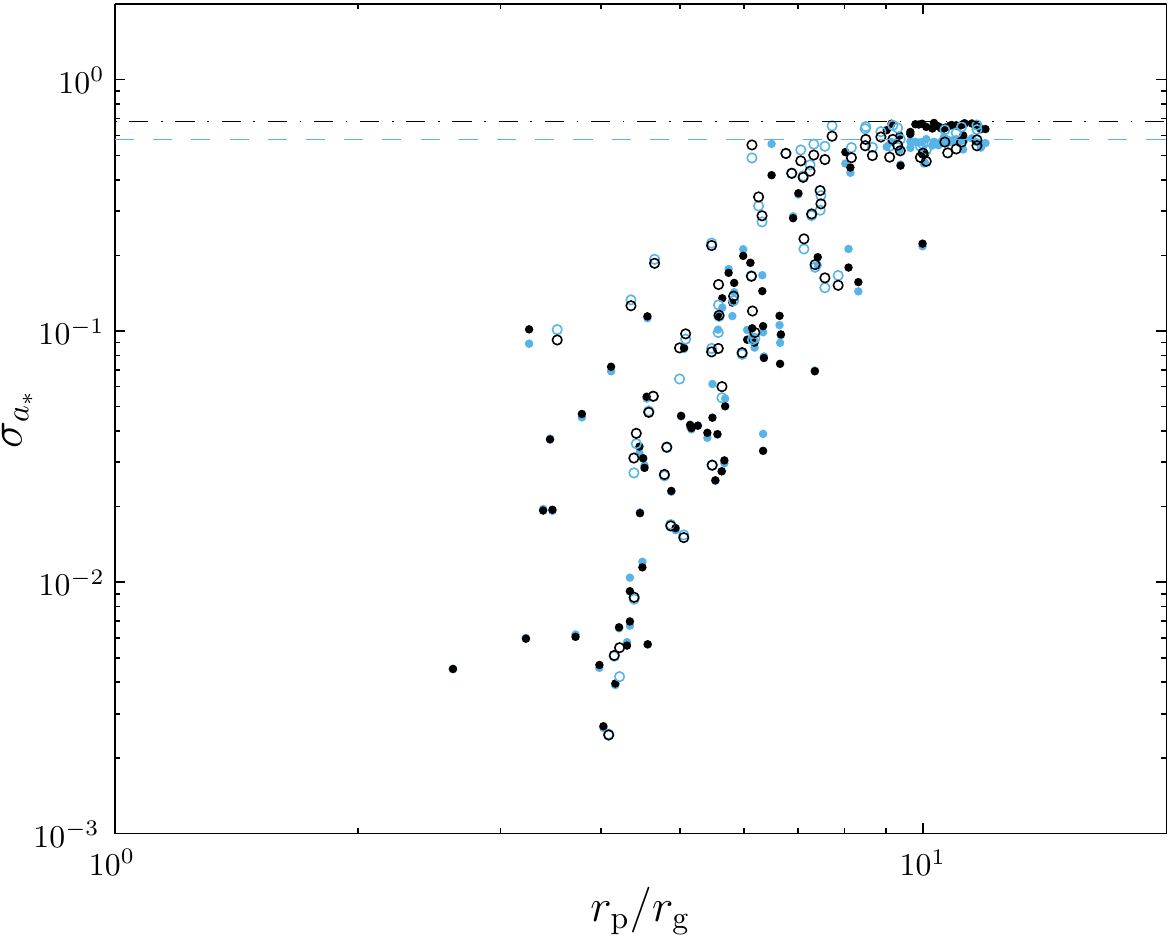}} \\
\caption{Distribution widths as functions of periapsis $r\sub{p}$ for M32. The light blue points are used for the standard deviation, black for the $68$-percentile half-width. The filled circles are converged runs and the open circles for those yet to converge. The dotted line is the current uncertainty for $M$. The dashed line is the standard deviation for a uniform $a_\ast$ distribution and the dot--dashed line is the equivalent $68$-percentile half-width.\label{fig:sigmas-M32}}
\end{center}
\end{figure*}

\begin{figure*}
\begin{center}
\subfigure[Natural logarithm of MBH mass.]{\includegraphics[width=0.43\textwidth]{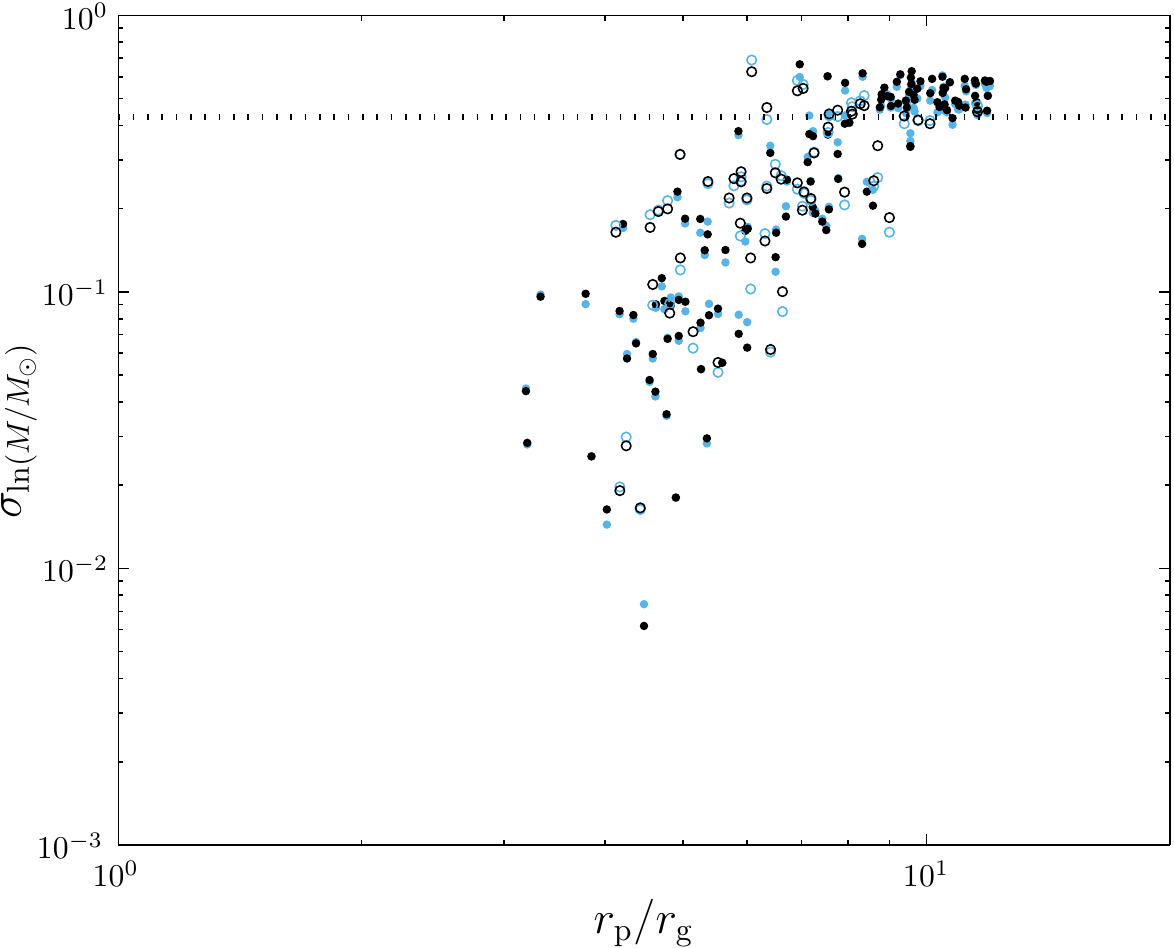}} \quad
\subfigure[Dimensionless spin.]{\includegraphics[width=0.43\textwidth]{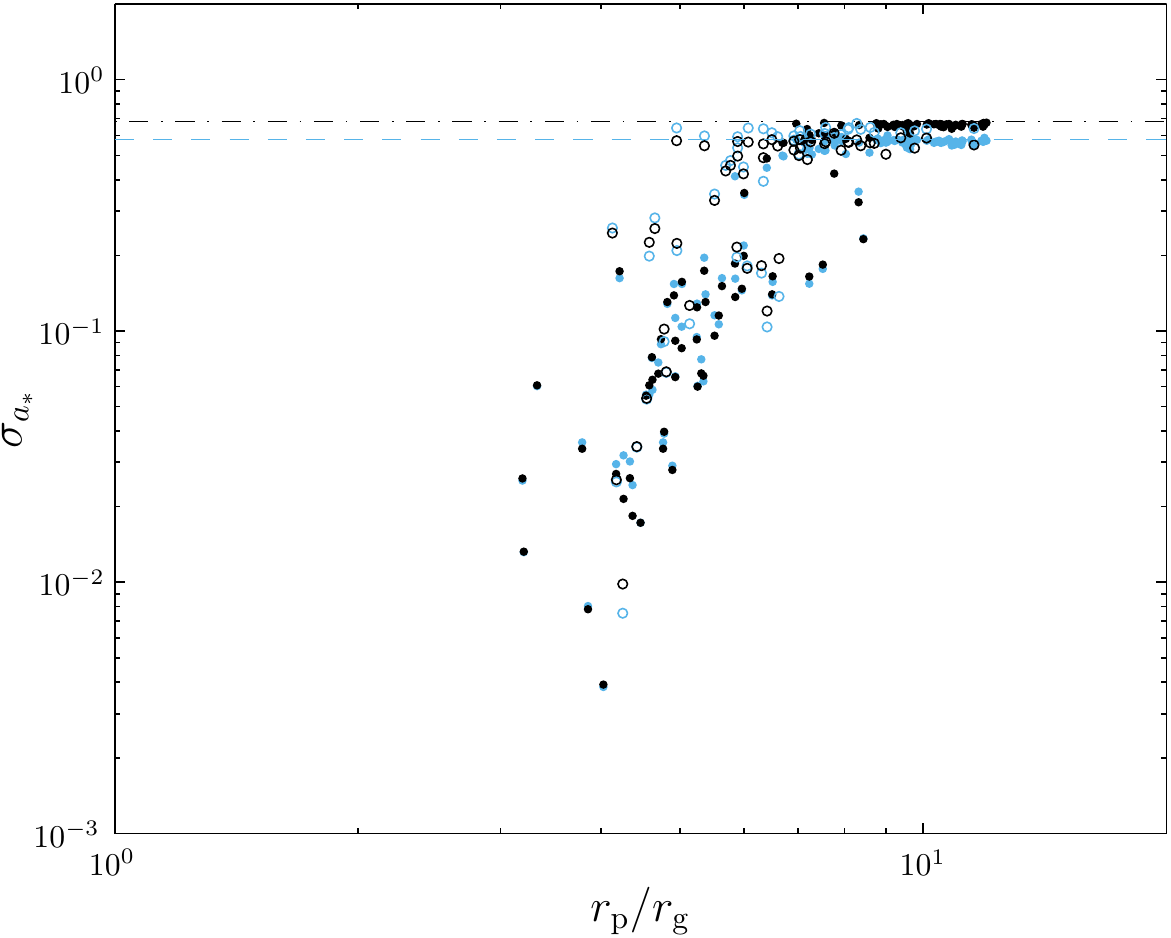}} \\
\caption{Distribution widths as functions of periapsis $r\sub{p}$ for NGC 4945. Conventions are identical to those in \figref{sigmas-M32}.\label{fig:sigmas-N4945}}
\end{center}
\end{figure*}

\begin{figure*}
\begin{center}
\subfigure[Natural logarithm of MBH mass.]{\includegraphics[width=0.43\textwidth]{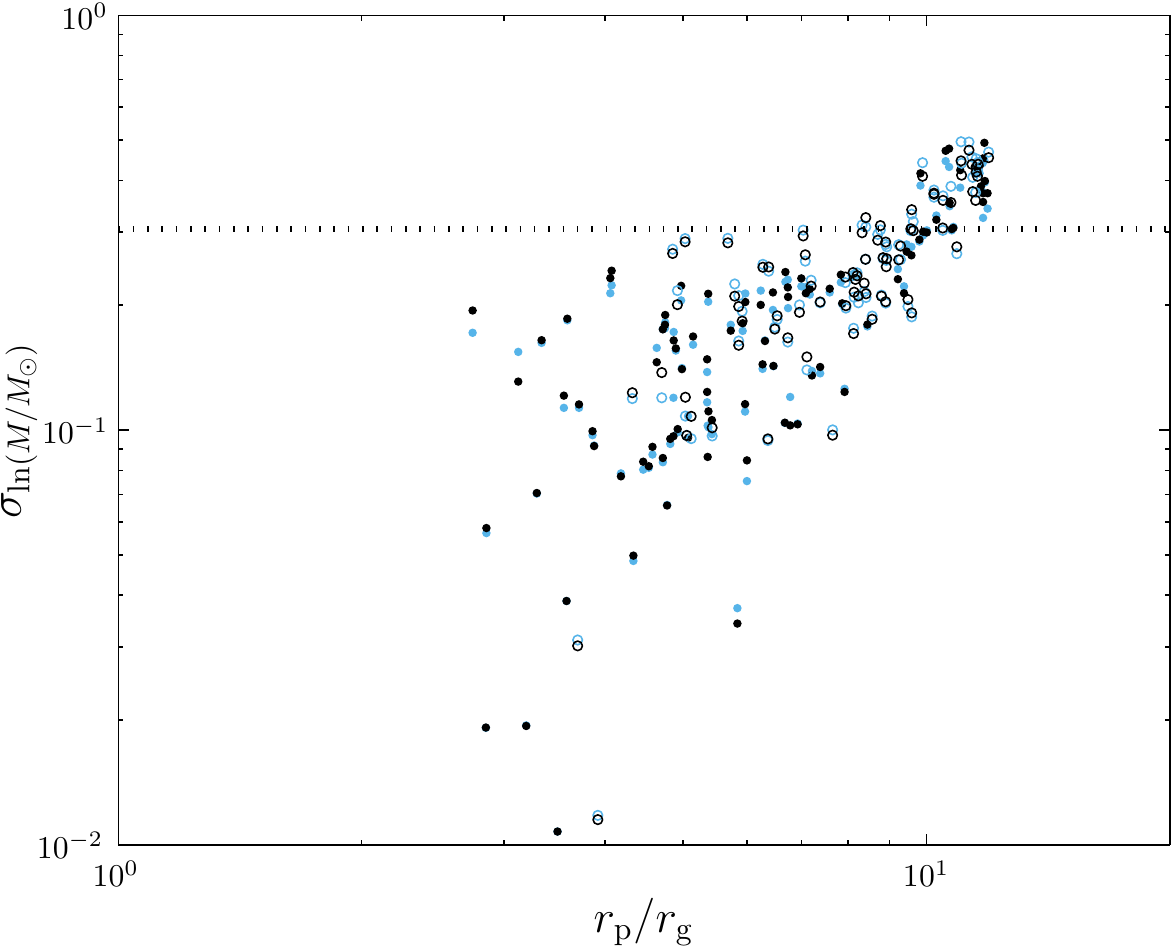}} \quad
\subfigure[Dimensionless spin.]{\includegraphics[width=0.43\textwidth]{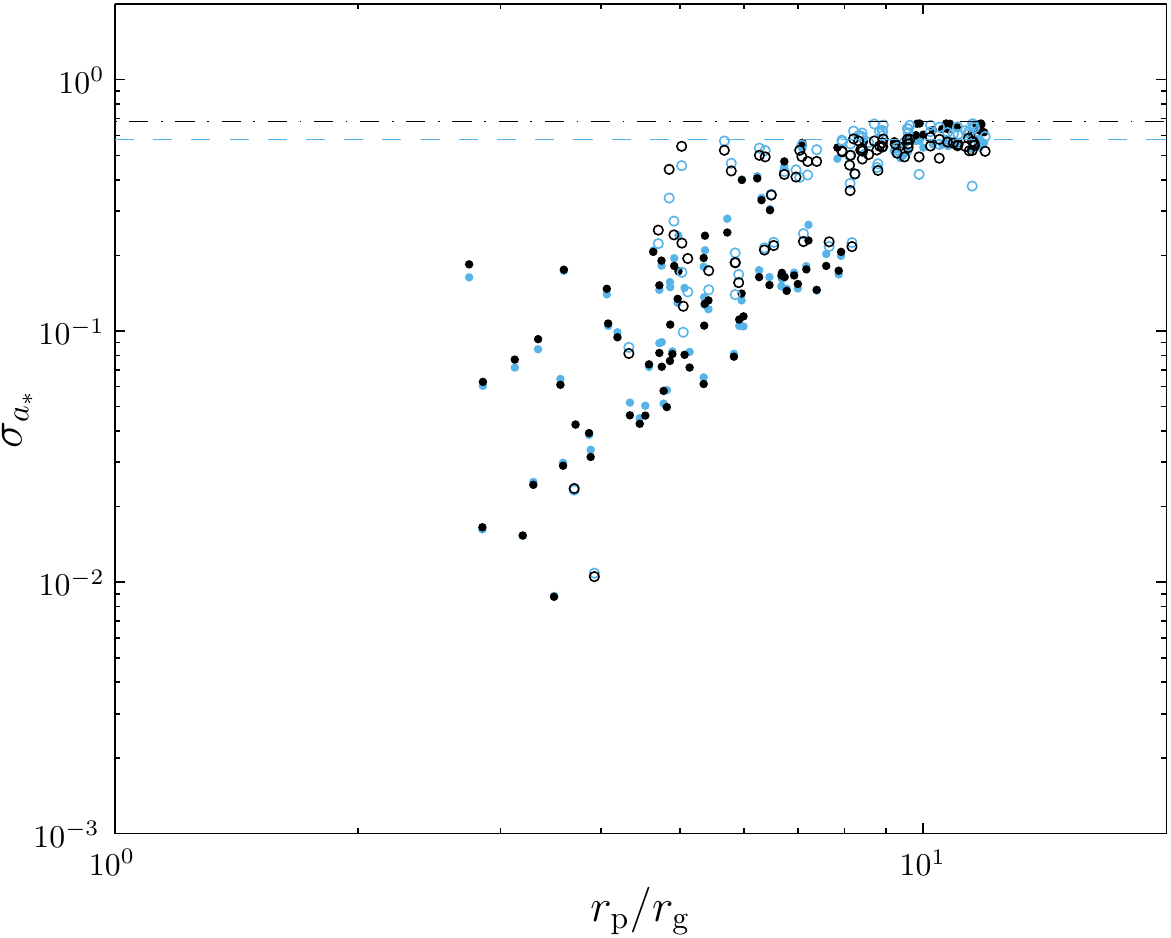}} \\
\caption{Distribution widths as functions of periapsis $r\sub{p}$ for NGC 4395. Conventions are identical to those in \figref{sigmas-M32}.\label{fig:sigmas-N4395}}
\end{center}
\end{figure*}

Characteristic distribution widths for the (logarithm of the) MBH mass and the spin are shown in \figref{sigmas-M32}, \ref{fig:sigmas-N4945} and \ref{fig:sigmas-N4395} for M32, NGC 4945 and NGC 4395 respectively. Plotted are the standard deviation $\sigma\sub{SD}$ and the half-width of the $p = 0.68$ range calculated from the $k$-d tree $\sigma_{0.68}$. These widths are equal for a normal distribution. The filled circles are used for runs that appear to have converged. The open circles are for those yet to converge, but which appear to be approaching an equilibrium state; widths should be accurate to within $\sim 10\%$.

\begin{figure}
\begin{center}
   \includegraphics[width=0.43\textwidth]{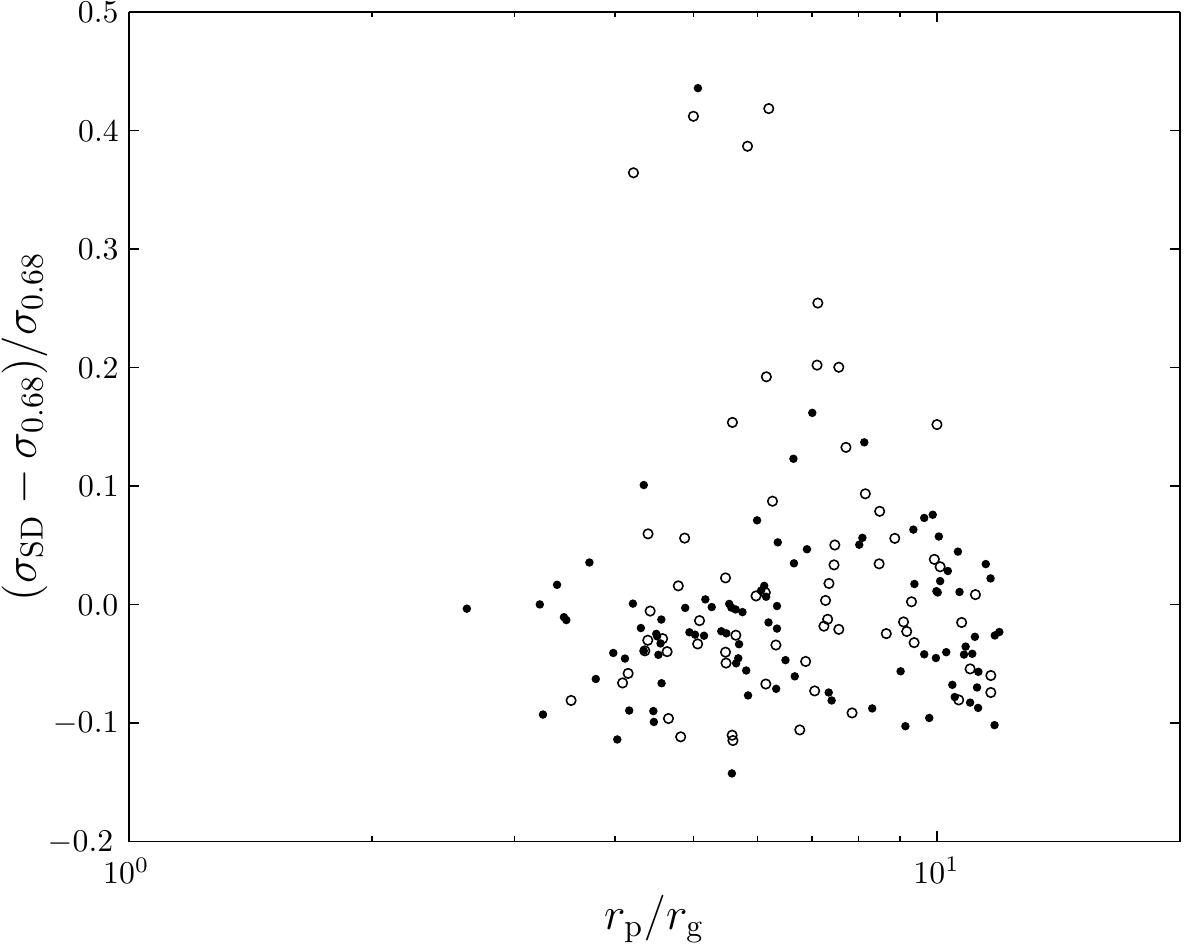}
\caption{Fractional difference between the standard deviation $\sigma\sub{SD}$ and the $68$-percentile half-width $\sigma_{0.68}$ of the marginalised posterior for $\ln(M/M_\odot)$ as a function of periapsis for bursts from M32. The filled circles are converged runs nad the open circles for those yet to converge.\label{fig:sigma-ratio-mass}}
\end{center}
\end{figure}

The widths, corresponding to potential parameter accuracies, improve rapidly with decreasing periapsis. The two widths, $\sigma\sub{SD}$ and $\sigma_{0.68}$, are typically of similar sizes, despite manifest non-Gaussianity. This is true for all parameters: the greatest differences are when the distributions are strongly multimodal. \Figref{sigma-ratio-mass} shows the fractional difference between the two widths of the $\ln(M/M_\odot)$ distribution for EMRBs from M32.
The widths for $a_\ast$ are similar; the widths for $\phi\sub{p}$ show the greatest difference, where $\sigma\sub{SD}$ may be a factor of a few larger than $\sigma_{0.68}$. Both $\sigma\sub{SD}$ and $\sigma_{0.68}$ tend to the appropriate limits for uninformative distributions.

In the best case, uncertainties in mass and spin may be only one part in $10^2$. As might be expected from \figref{detect}, M32 has the smallest widths followed by NGC 4945 and then NGC 4395. The spin width saturates about the value expected from a uniform distribution. At this point, we can no longer constrain the spin. The transition does not show any clear correlation with the magnitude of the spin, but is predominantly determined by the periapsis and SNR.

The other parameters show similar behaviour. The angular variables also reach maximum widths, corresponding to uninformative distributions. This does not appear to be directly tied to the spin width.

Potentially, an EMRB could place useful constraints on the mass and spin of an MBH if the periapse radius is small enough.\footnote{Here, we assume that a mass measurement is useful if its accuracy is smaller than the current measurement uncertainty, and a spin measurement is useful if it provides any constraint.} For M32 we require $r\sub{p} \lesssim 8 r\sub{g}$; for NGC 4945 we require $r\sub{p} \lesssim 8 r\sub{g}$ and $r\sub{p} \lesssim 7 r\sub{g}$ for mass and spin measurements, respectively, and for NGC 4395 we require $r\sub{p} \lesssim 9 r\sub{g}$ and $r\sub{p} \lesssim 8 r\sub{g}$, respectively. Since the range of useful periapses is small, we expect useful EMRBs originating from any individual galaxy to be rare. However, because there are many galaxies hosting potential sources, the probability of seeing any useful EMRBs need not be negligible. Therefore, EMRBs could be a useful astronomical tool.

\section{Conclusions}\label{sec:End}

We have studied extreme-mass-ratio bursts from extragalactic sources. The SNR of EMRBs has a fundamental scaling with the system parameters. Removing these proportionalities gives a scaled SNR that can be specified as a function of the characteristic frequency $f_\ast$. Using these relations allows us to calculate the maximum distance to which EMRBs from a system containing an MBH of a given mass can be detected.

The MBH in our own Galaxy is by far the best source for bursts; however, it is also possible to detect bursts from extragalactic sources. We were previously too pessimistic about this possibility \citep{Berry2013}. In particular, M32 is a promising candidate. This is good news for any space-borne GW detectors, as EMRBs can be added to their list of potential sources.

Utilising the classic \textit{LISA} design, EMRBs from a $10 M_\odot$ orbiting CO could be detected out to a distance of $\sim 100\units{Mpc}$. With the descoped \textit{eLISA} design, this decreases to $\sim 10\units{Mpc}$. This may drastically reduce the chance of observing an EMRB. For both detectors, sensitivity is maximal for MBHs of $M \sim 10^6$--$10^7 M_\odot$, being at slightly higher masses for \textit{LISA} than for \textit{eLISA}. We can detect bursts from systems with high MBH spins out to greater distance; hence, the EMRB event rate would be enhanced if MBH spins naturally tend to higher values, perhaps as a consequence of accretion.

However, we must still be cautious: EMRBs may be rare and the event rate may prevent us from observing any over a realistic mission lifetime. Bursts from any given extragalactic source should be less common than from the Galactic Centre, although this may be slightly ameliorated by the larger number of galaxies hosting potential source systems. In a companion paper \citep{Berry2013b}, we shall look at the expectations for bursts from our own Galaxy.\footnote{In this we build a model carefully expanding the early work of \citet*{Hopman2007}. The overall rates are similar to those presented in \citet{Hopman2007}.} Adapting results from this by picking a maximum periapse radius of $8 r\sub{g}$, which is approximately the point at which bursts from our sample become interesting, we calculate a rough event rate of $\sim 0.2\units{yr^{-1}}$. This is an order of magnitude estimate using parameters for the Galaxy, which are much better constrained than for other galaxies. Using this as a reference, it appears that extragalactic EMRBs would not be common. However, assuming a $2\units{yr}$ mission lifetime and that our extragalactic candidates add up to be equivalent to a few Milky Ways, it might not be surprising to observe one or more extragalactic EMRBs. 

Extragalactic EMRBs can provide good measurements of MBH mass and spin, but only across an extremely narrow range of periapses. We studied M32, NGC 4945 and NGC 4395 as examples. For all three we found that it is possible to extract information from bursts. The uncertainty may be one part in $10^2$--$10^3$ for M32, and slightly worse for NGC 4945 and NGC 4395, at about one part in $10^2$. These are not as good as the constraints from Galactic EMRBs, where the uncertainties could be as small as one part in $10^4$ \citep{Berry2013}, but would still be of great astrophysical interest. These extragalactic MBHs are much harder to study than the MBH in our own Galaxy and we have not yet been able to measure a spin value even for that MBH. Any measurement of spin would give us a unique glimpse into the formation history of the host galaxy.

These results have been obtained assuming the classic \textit{LISA} design. The first millihertz space-borne interferometer is likely to have a descoped design such as the proposed \textit{eLISA}. This concept could be revised in the near future and so we have not used it to produce results. The effect of the reduced sensitivity would be to reduce the SNR and increase the widths of the posterior distributions. We expect the trends in \figref{sigmas-M32}, \ref{fig:sigmas-N4945} and \ref{fig:sigmas-N4395} to move upwards and saturate at smaller periapses.

Extreme-mass-ratio bursts could be used to place useful constraints on the mass and spin of a nearby MBH if the periapse radius is small enough. Considering the promising candidates of M32, NGC 4945 and NGC 4395, we find that $r\sub{p} \lesssim 8 r\sub{g}$ typically gives insightful constraints. Such orbits are likely to be rare, but just a single such burst from any of the potential galaxies could give us information that is otherwise inaccessible. This is a tantalising prospect.

\section*{Acknowledgements}

The authors would like to thank Donald Lynden-Bell for useful suggestions, and Ilya Mandel and Trevor Sidery for interesting conversations on $k$-d trees. CPLB is supported by STFC. JRG is supported by the Royal Society. The MCMC simulations were performed using the Darwin Supercomputer of the University of Cambridge High Performance Computing Service (\url{http://www.hpc.cam.ac.uk/}), provided by Dell Inc.\ using Strategic Research Infrastructure Funding from the Higher Education Funding Council for England.

\bibliographystyle{mn3e}
\bibliography{Extragalactic}

\begin{thebibliography}{76}
\expandafter\ifx\csname natexlab\endcsname\relax\def\natexlab#1{#1}\fi

\bibitem[{Adams(1979)}]{Adams1979}
Adams D., 1979, {The Hitchikers' Guide to the Galaxy}, 1st edn. Pan Books,
  London

\bibitem[{Alexander \& Hopman(2009)}]{Alexander2009}
Alexander T., Hopman C., 2009, ApJ, 697, 1861

\bibitem[{Amaro-Seoane {et~al}\mbox{.}(2012)Amaro-Seoane, Aoudia, Babak,
  Bin{\'e}truy, Berti, Boh{\'e}, Caprini, Colpi, Cornish, Danzmann, Dufaux,
  Gair, Jennrich, Jetzer, Klein, Lang, Lobo, Littenberg, McWilliams, Nelemans,
  Petiteau, Porter, Schutz, Sesana, Stebbins, Sumner, Vallisneri, Vitale,
  Volonteri, \& Ward}]{Amaro-Seoane2012a}
Amaro-Seoane P. {et~al.}, 2012, Class.\ Quantum Grav., 29, 124016

\bibitem[{Amaro-Seoane {et~al}\mbox{.}(2007)Amaro-Seoane, Gair, Freitag,
  Miller, Mandel, Cutler, \& Babak}]{Amaro-Seoane2007}
Amaro-Seoane P., Gair J.~R., Freitag M., Miller M.~C., Mandel I., Cutler C.~J.,
  Babak S., 2007, Class.\ Quantum Grav., 24, R113

\bibitem[{Antonucci {et~al}\mbox{.}(2012)Antonucci, Armano, Audley, Auger,
  Benedetti, Binetruy, Bogenstahl, Bortoluzzi, Bosetti, Brandt, Caleno,
  Ca\~{n}izares, Cavalleri, Cesa, Chmeissani, Conchillo, Congedo, Cristofolini,
  Cruise, Danzmann, {De Marchi}, Diaz-Aguilo, Diepholz, Dixon, Dolesi, Dunbar,
  Fauste, Ferraioli, Ferrone, Fichter, Fitzsimons, Freschi, Marin,
  Marirrodriga, Gerndt, Gesa, Gilbert, Giardini, Grimani, Grynagier, Guillaume,
  Guzm\'{a}n, Harrison, Heinzel, Hern\'{a}ndez, Hewitson, Hollington, Hough,
  Hoyland, Hueller, Huesler, Jennrich, Jetzer, Johlander, Karnesis, Killow,
  Llamas, Lloro, Lobo, Maarschalkerweerd, Madden, Mance, Mateos, McNamara,
  Mendes, Mitchell, Monsky, Nicolini, Nicolodi, Nofrarias, Pedersen,
  Perreur-Lloyd, Plagnol, Prat, Racca, Ramos-Castro, Reiche, Perez, Robertson,
  Rozemeijer, Sanjuan, Schleicher, Schulte, Shaul, Stagnaro, Strandmoe, Steier,
  Sumner, Taylor, Texier, Trenkel, Tu, Vitale, Wanner, Ward, Waschke, Wass,
  Weber, Ziegler, \& Zweifel}]{Antonucci2012}
Antonucci F. {et~al.}, 2012, Class.\ Quantum Grav., 29, 124014

\bibitem[{Anza {et~al}\mbox{.}(2005)Anza, Armano, Balaguer, Benedetti,
  Boatella, Bosetti, Bortoluzzi, Brandt, Braxmaier, Caldwell, Carbone,
  Cavalleri, Ciccolella, Cristofolini, Cruise, Lio, Danzmann, Desiderio,
  Dolesi, Dunbar, Fichter, Garcia, Garcia-Berro, Marin, Gerndt, Gianolio,
  Giardini, Gruenagel, Hammesfahr, Heinzel, Hough, Hoyland, Hueller, Jennrich,
  Johann, Kemble, Killow, Kolbe, Landgraf, Lobo, Lorizzo, Mance, Middleton,
  Nappo, Nofrarias, Racca, Ramos, Robertson, Sallusti, Sandford, Sanjuan,
  Sarra, Selig, Shaul, Smart, Smit, Stagnaro, Sumner, Tirabassi, Tobin, Vitale,
  Wand, Ward, Weber, \& Zweifel}]{Anza2005}
Anza S. {et~al.}, 2005, Class.\ Quantum Grav., 22, S125

\bibitem[{Babak {et~al}\mbox{.}(2007)Babak, Fang, Gair, Glampedakis, \&
  Hughes}]{Babak2007}
Babak S., Fang H., Gair J., Glampedakis K., Hughes S., 2007, Phys.\ Rev.\ D,
  75, 024005

\bibitem[{Bahcall \& Wolf(1977)}]{Bahcall1977}
Bahcall J.~N., Wolf R.~A., 1977, ApJ, 216, 883

\bibitem[{Barack \& Cutler(2004)}]{Barack2004}
Barack L., Cutler C., 2004, Phys.\ Rev.\ D, 69, 082005

\bibitem[{Barausse(2012)}]{Barausse2012}
Barausse E., 2012, MNRAS, 423, 2533

\bibitem[{Bekenstein(1973)}]{Bekenstein1973}
Bekenstein J.~D., 1973, ApJ, 183, 657

\bibitem[{Bender {et~al}\mbox{.}(1998)Bender, Brillet, Ciufolini, Cruise,
  Cutler, Danzmann, Fidecaro, Folkner, Hough, McNamara, Peterseim, Robertson,
  Rodrigues, R\"{u}diger, Sandford, Sch\"{a}fer, Schilling, Schutz, Speake,
  Stebbins, Sumner, Touboul, Vinet, Vitale, Ward, \& Winkler}]{Bender1998}
Bender P. {et~al.}, 1998, {LISA Pre-Phase A Report}. Tech. rep.,
  Max-Planck-Institut f\"{u}r Quantenoptik, Garching

\bibitem[{Bender {et~al}\mbox{.}(2005)Bender, Kormendy, Bower, Green, Thomas,
  Danks, Gull, Hutchings, Joseph, Kaiser, Lauer, Nelson, Richstone, Weistrop,
  \& Woodgate}]{Bender2005}
Bender R. {et~al.}, 2005, ApJ, 631, 280

\bibitem[{Berry \& Gair(2013{\natexlab{a}})}]{Berry2013b}
Berry C. P.~L., Gair J.~R., 2013{\natexlab{a}}, MNRAS, 435, 3521

\bibitem[{Berry \& Gair(2013{\natexlab{b}})}]{Berry2013}
Berry C. P.~L., Gair J.~R., 2013{\natexlab{b}}, MNRAS, 429, 589

\bibitem[{Berti \& Volonteri(2008)}]{Berti2008}
Berti E., Volonteri M., 2008, ApJ, 684, 822

\bibitem[{Brenneman {et~al}\mbox{.}(2011)Brenneman, Reynolds, Nowak, Reis,
  Trippe, Fabian, Iwasawa, Lee, Miller, Mushotzky, Nandra, \&
  Volonteri}]{Brenneman2011}
Brenneman L.~W. {et~al.}, 2011, ApJ, 736, 103

\bibitem[{Chandrasekhar(1998)}]{Chandrasekhar1998}
Chandrasekhar S., 1998, {The Mathematical Theory of Black Holes}, Oxford
  Classic Texts in the Physical Sciences. Oxford Univ.\ Press, Oxford

\bibitem[{Cutler(1998)}]{Cutler1998}
Cutler C., 1998, Phys.\ Rev.\ D, 57, 7089

\bibitem[{Cutler \& Flanagan(1994)}]{Cutler1994}
Cutler C., Flanagan E.~E., 1994, Phys.\ Rev.\ D, 49, 2658

\bibitem[{Danzmann \& R\"{u}diger(2003)}]{Danzmann2003}
Danzmann K., R\"{u}diger A., 2003, Class.\ Quantum Grav., 20, S1

\bibitem[{de~Berg {et~al}\mbox{.}(2008)de~Berg, Cheong, van Kreveld, \&
  Overmars}]{Berg2008}
de~Berg M., Cheong O., van Kreveld M., Overmars M., 2008, {Computational
  Geometry: Algorithms and Applications}, 3rd edn. Springer, Berlin, p. 386

\bibitem[{Dotti {et~al}\mbox{.}(2013)Dotti, Colpi, Pallini, Perego, \&
  Volonteri}]{Dotti2013}
Dotti M., Colpi M., Pallini S., Perego A., Volonteri M., 2013, ApJ, 762, 68

\bibitem[{Drasco \& Hughes(2004)}]{Drasco2004}
Drasco S., Hughes S., 2004, Phys.\ Rev.\ D, 69, 044015

\bibitem[{Ferrarese \& Ford(2005)}]{Ferrarese2005}
Ferrarese L., Ford H., 2005, Space Sci.\ Rev., 116, 523

\bibitem[{Finn(1992)}]{Finn1992}
Finn L.~S., 1992, Phys.\ Rev.\ D, 46, 5236

\bibitem[{Gair {et~al}\mbox{.}(2005)Gair, Kennefick, \& Larson}]{Gair2005}
Gair J.~R., Kennefick D.~J., Larson S.~L., 2005, Phys.\ Rev.\ D, 72, 084009

\bibitem[{Gammie {et~al}\mbox{.}(2004)Gammie, Shapiro, \&
  McKinney}]{Gammie2004}
Gammie C.~F., Shapiro S.~L., McKinney J.~C., 2004, ApJ, 602, 312

\bibitem[{Gillessen {et~al}\mbox{.}(2009)Gillessen, Eisenhauer, Trippe,
  Alexander, Genzel, Martins, \& Ott}]{Gillessen2009}
Gillessen S., Eisenhauer F., Trippe S., Alexander T., Genzel R., Martins F.,
  Ott T., 2009, ApJ, 692, 1075

\bibitem[{Glampedakis {et~al}\mbox{.}(2002)Glampedakis, Hughes, \&
  Kennefick}]{Glampedakis2002}
Glampedakis K., Hughes S., Kennefick D., 2002, Phys.\ Rev.\ D, 66, 064005

\bibitem[{Glampedakis \& Kennefick(2002)}]{Glampedakis2002a}
Glampedakis K., Kennefick D., 2002, Phys.\ Rev.\ D, 66, 044002

\bibitem[{Gonz{\'a}lez {et~al}\mbox{.}(2007)Gonz{\'a}lez, Sperhake,
  Br{\"u}gmann, Hannam, \& Husa}]{Gonzalez2007}
Gonz{\'a}lez J.~A., Sperhake U., Br{\"u}gmann B., Hannam M., Husa S., 2007,
  Phys.\ Rev.\ Lett., 98, 091101

\bibitem[{Graham(2008)}]{Graham2008}
Graham A.~W., 2008, Publ.\ Astron.\ Soc.\ Aust., 25, 167

\bibitem[{Graham {et~al}\mbox{.}(2011)Graham, Onken, Athanassoula, \&
  Combes}]{Graham2011}
Graham A.~W., Onken C.~A., Athanassoula E., Combes F., 2011, MNRAS, 412, 2211

\bibitem[{Graham \& Scott(2013)}]{Graham2013}
Graham A.~W., Scott N., 2013, ApJ, 764, 151

\bibitem[{Greenhill {et~al}\mbox{.}(2003)Greenhill, Booth, Ellingsen,
  Herrnstein, Jauncey, McCulloch, Moran, Norris, Reynolds, \&
  Tzioumis}]{Greenhill2003}
Greenhill L.~J. {et~al.}, 2003, ApJ, 590, 162

\bibitem[{Greenhill {et~al}\mbox{.}(1997)Greenhill, Moran, \&
  Herrnstein}]{Greenhill1997}
Greenhill L.~J., Moran J.~M., Herrnstein J.~R., 1997, ApJ, 481, L23

\bibitem[{Haario {et~al}\mbox{.}(1999)Haario, Saksman, \&
  Tamminen}]{Haario1999}
Haario H., Saksman E., Tamminen J., 1999, Comput.\ Stat., 14, 375

\bibitem[{Hastings(1970)}]{Hastings1970}
Hastings W.~K., 1970, Biometrika, 57, 97

\bibitem[{Hobson {et~al}\mbox{.}(2006)Hobson, Efstathiou, \&
  Lasenby}]{Hobson2006}
Hobson M.~P., Efstathiou G., Lasenby A., 2006, {General Relativity: An
  Introduction for Physicists}. Cambridge University Press, Cambridge

\bibitem[{Hopman {et~al}\mbox{.}(2007)Hopman, Freitag, \& Larson}]{Hopman2007}
Hopman C., Freitag M., Larson S.~L., 2007, MNRAS, 378, 129

\bibitem[{Hughes \& Blandford(2003)}]{Hughes2003}
Hughes S.~A., Blandford R.~D., 2003, ApJ, 585, L101

\bibitem[{Jaranowski \& Kr{\'o}lak(2012)}]{Jaranowski2012}
Jaranowski P., Kr{\'o}lak A., 2012, Living Rev.\ Relativ., 15

\bibitem[{Jennrich {et~al}\mbox{.}(2011)Jennrich, Binetruy, Colpi, Danzmann,
  Jetzer, Lobo, Nelemans, Schutz, Stebbins, Sumner, Vitale, \&
  Ward}]{Jennrich2011}
Jennrich O. {et~al.}, 2011, {NGO Revealing a Hidden Universe: Opening a New
  Chapter of Discovery}. Tech. rep., European Space Agency, Noordwijk

\bibitem[{Karachentsev {et~al}\mbox{.}(2004)Karachentsev, Karachentseva,
  Huchtmeier, \& Makarov}]{Karachentsev2004}
Karachentsev I.~D., Karachentseva V.~E., Huchtmeier W.~K., Makarov D.~I., 2004,
  AJ, 127, 2031

\bibitem[{Karachentsev {et~al}\mbox{.}(2007)Karachentsev, Tully, Dolphin,
  Sharina, Makarova, Makarov, Sakai, Shaya, Kashibadze, Karachentseva, \&
  Rizzi}]{Karachentsev2007}
Karachentsev I.~D. {et~al.}, 2007, AJ, 133, 504

\bibitem[{King \& Pringle(2006)}]{King2006}
King A.~R., Pringle J.~E., 2006, MNRAS, 373, L90

\bibitem[{King {et~al}\mbox{.}(2008)King, Pringle, \& Hofmann}]{King2008}
King A.~R., Pringle J.~E., Hofmann J.~A., 2008, MNRAS, 385, 1621

\bibitem[{Kormendy \& Richstone(1995)}]{Kormendy1995}
Kormendy J., Richstone D., 1995, ARA\&A, 33, 581

\bibitem[{Lynden-Bell \& Rees(1971)}]{Lynden-Bell1971}
Lynden-Bell D., Rees M.~J., 1971, MNRAS, 152, 461

\bibitem[{MacKay(2003)}]{MacKay2003}
MacKay D. J.~C., 2003, {Information Theory, Inference and Learning Algorithms}.
  Cambridge Univ.\ Press, Cambridge, p. 640

\bibitem[{Magorrian {et~al}\mbox{.}(1998)Magorrian, Tremaine, Richstone,
  Bender, Bower, Dressler, Faber, Gebhardt, Green, Grillmair, Kormendy, \&
  Lauer}]{Magorrian1998}
Magorrian J. {et~al.}, 1998, AJ, 115, 2285

\bibitem[{McNamara {et~al}\mbox{.}(2013)McNamara, Antonucci, Armano, Audley,
  Auger, Benedetti, Binetruy, Bogenstahl, Bortoluzzi, Brandt, Caleno,
  Cavalleri, Congedo, Cruise, Danzmann, {De Marchi}, Diaz-Aguilo, Diepholz,
  Dixton, Dolesi, Dumbar, Fauste, Ferraioli, Ferroni, Fichter, Fitzsimons,
  Freschi, {Garc\'{\i}a Marirrodriga}, Gerndt, Gesa, Gibert, Giardini, Grimani,
  Grynagier, Guzm\'{a}n, Harrison, Heinzel, Hewitson, Hollington, Hoyland,
  Hueller, Huesler, Jennrich, Jetzer, Johlander, Karnesis, Korsakova, Killow,
  Llamas, Lloro, Lobo, Maarschalkerweerd, Madden, Mance, Martin, Mateos,
  Mendes, Mitchell, Nicolodi, Nofrarias, Perreur-Lloyd, Plagnol, Prat,
  Ramos-Castro, Reiche, {Romera Perez}, Robertson, Rozemeijer, Russano,
  Schleicher, Shaul, Sopuerta, Sumner, Taylor, Texier, Trenkel, Tu, Vitale,
  Wanner, Ward, Waschke, Wass, Wealthy, Wen, Weber, Ziegler, \&
  Zweifel}]{McNamara2013}
McNamara P. {et~al.}, 2013, in ASP Conf.\ Ser., Auger G., Bin{\'e}truy P.,
  Plagnol E., eds., Vol. 467, Astron.\ Soc.\ Pac., San Francisco, pp. 5--16

\bibitem[{Metropolis {et~al}\mbox{.}(1953)Metropolis, Rosenbluth, Rosenbluth,
  Teller, \& Teller}]{Metropolis1953}
Metropolis N., Rosenbluth A.~W., Rosenbluth M.~N., Teller A.~H., Teller E.,
  1953, J.\ Chem.\ Phys., 21, 1087

\bibitem[{Misner {et~al}\mbox{.}(1973)Misner, Thorne, \& Wheeler}]{Misner1973}
Misner C.~W., Thorne K.~S., Wheeler J.~A., 1973, {Gravitation}. W. H. Freeman,
  New York

\bibitem[{Nowak {et~al}\mbox{.}(2010)Nowak, Thomas, Erwin, Saglia, Bender, \&
  Davies}]{Nowak2010}
Nowak N., Thomas J., Erwin P., Saglia R.~P., Bender R., Davies R.~I., 2010,
  MNRAS, 403, 646

\bibitem[{Peterson {et~al}\mbox{.}(2005)Peterson, Bentz, Desroches, Filippenko,
  Ho, Kaspi, Laor, Maoz, Moran, Pogge, \& Quillen}]{Peterson2005}
Peterson B.~M. {et~al.}, 2005, ApJ, 632, 799

\bibitem[{Press(1977)}]{Press1977}
Press W., 1977, Phys.\ Rev.\ D, 15, 965

\bibitem[{Preto \& Amaro-Seoane(2010)}]{Preto2010}
Preto M., Amaro-Seoane P., 2010, ApJ, 708, L42

\bibitem[{Rekola {et~al}\mbox{.}(2005)Rekola, Richer, McCall, Valtonen,
  Kotilainen, \& Flynn}]{Rekola2005}
Rekola R., Richer M.~G., McCall M.~L., Valtonen M.~J., Kotilainen J.~K., Flynn
  C., 2005, MNRAS, 361, 330

\bibitem[{Rodr{\'i}guez-Rico {et~al}\mbox{.}(2006)Rodr{\'i}guez-Rico, Goss,
  Zhao, Gomez, \& Anantharamaiah}]{Rodriguez-Rico2006}
Rodr{\'i}guez-Rico C.~A., Goss W.~M., Zhao J.-H., Gomez Y., Anantharamaiah
  K.~R., 2006, ApJ, 644, 914

\bibitem[{Rubbo {et~al}\mbox{.}(2006)Rubbo, Holley-Bockelmann, \&
  Finn}]{Rubbo2006}
Rubbo L.~J., Holley-Bockelmann K., Finn L.~S., 2006, ApJ, 649, L25

\bibitem[{Ruffini \& Sasaki(1981)}]{Ruffini1981}
Ruffini R., Sasaki M., 1981, Prog.\ Theor.\ Phys., 66, 1627

\bibitem[{Sidery {et~al}\mbox{.}(2013)Sidery, Gair, \& Mandel}]{Sidery2013}
Sidery T., Gair J.~R., Mandel I., 2013, in preparation

\bibitem[{So{\l}tan(1982)}]{Soltan1982}
So{\l}tan A., 1982, MNRAS, 200, 115

\bibitem[{Tanaka {et~al}\mbox{.}(1993)Tanaka, Shibata, Sasaki, Tagoshi, \&
  Nakamura}]{Tanaka1993}
Tanaka T., Shibata M., Sasaki M., Tagoshi H., Nakamura T., 1993, Prog.\ Theor.\
  Phys., 90, 65

\bibitem[{Thim {et~al}\mbox{.}(2004)Thim, Hoessel, Saha, Claver, Dolphin, \&
  Tammann}]{Thim2004}
Thim F., Hoessel J.~G., Saha A., Claver J., Dolphin A., Tammann G.~A., 2004,
  AJ, 127, 2322

\bibitem[{Tonry {et~al}\mbox{.}(2001)Tonry, Dressler, Blakeslee, Ajhar,
  Fletcher, Luppino, Metzger, \& Moore}]{Tonry2001}
Tonry J.~L., Dressler A., Blakeslee J.~P., Ajhar E.~A., Fletcher A.~B., Luppino
  G.~A., Metzger M.~R., Moore C.~B., 2001, ApJ, 546, 681

\bibitem[{Verolme {et~al}\mbox{.}(2002)Verolme, Cappellari, Copin, van~der
  Marel, Bacon, Bureau, Davies, Miller, \& de~Zeeuw}]{Verolme2002}
Verolme E.~K. {et~al.}, 2002, MNRAS, 335, 517

\bibitem[{Volonteri(2010)}]{Volonteri2010}
Volonteri M., 2010, The Astronomy and Astrophysics Review, 18, 279

\bibitem[{Volonteri {et~al}\mbox{.}(2005)Volonteri, Madau, Quataert, \&
  Rees}]{Volonteri2005}
Volonteri M., Madau P., Quataert E., Rees M.~J., 2005, ApJ, 620, 69

\bibitem[{Volonteri {et~al}\mbox{.}(2012)Volonteri, Sikora, Lasota, \&
  Merloni}]{Volonteri2012a}
Volonteri M., Sikora M., Lasota J.-P., Merloni A., 2012, {The evolution of
  active galactic nuclei and their spins}, arXiv:1210.1025 [astro-ph.HE]

\bibitem[{Walton {et~al}\mbox{.}(2013)Walton, Nardini, Fabian, Gallo, \&
  Reis}]{Walton2013}
Walton D.~J., Nardini E., Fabian A.~C., Gallo L.~C., Reis R.~C., 2013, MNRAS,
  428, 2901

\bibitem[{Weinberg(2012)}]{Weinberg2012}
Weinberg M.~D., 2012, Bayesian Analysis, 7, 737

\bibitem[{Yu \& Tremaine(2002)}]{Yu2002}
Yu Q., Tremaine S., 2002, MNRAS, 335, 965

\bibitem[{Yunes {et~al}\mbox{.}(2008)Yunes, Sopuerta, Rubbo, \&
  Holley-Bockelmann}]{Yunes2008}
Yunes N.~N., Sopuerta C.~F., Rubbo L.~J., Holley-Bockelmann K., 2008, ApJ, 675,
  604

\end{thebibliography}

\appendix

\section{Example posterior distributions}\label{ap:posterior}

\begin{figure*}
\begin{center}
   \includegraphics[width=0.92\textwidth]{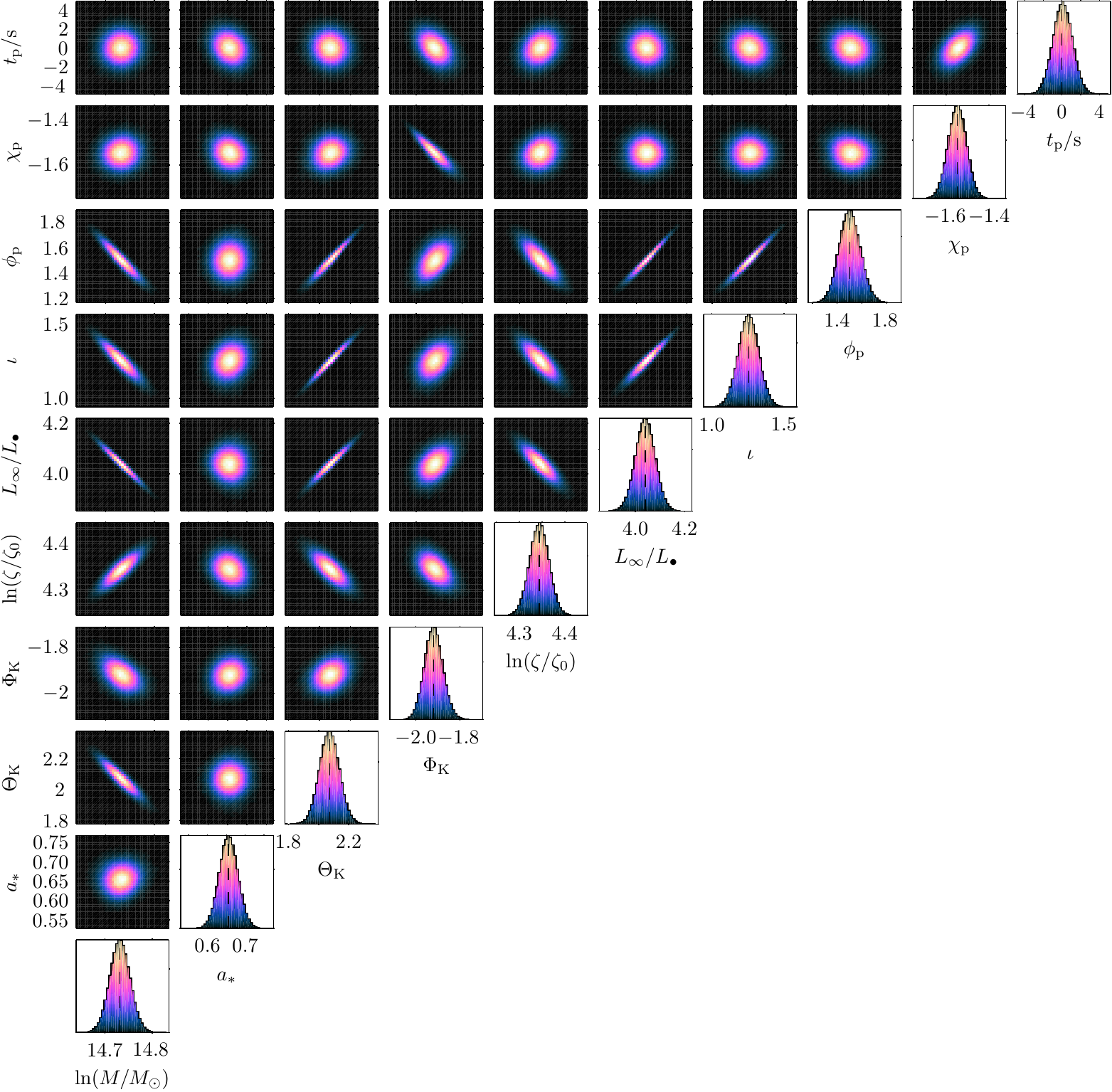}
\caption{Marginalised one- and two-dimensional posteriors (on the diagonal and above, respectively). The scales are identical in both types of plots. The dotted line indicates the true value. These distributions are exceptionally cromulent and well converged. Angular momentum is in units of $L_\bullet = GM c^{-1}$ and the scaled distance is in units of $\zeta_0 = 1 M_\odot^{-1}\units{kpc}$. The EMRB is from M32 and has $r\sub{p} \simeq 5.53 r\sub{g}$.\label{fig:MCMC-a}}
\end{center}
\end{figure*}

\begin{figure*}
\begin{center}
   \includegraphics[width=0.92\textwidth]{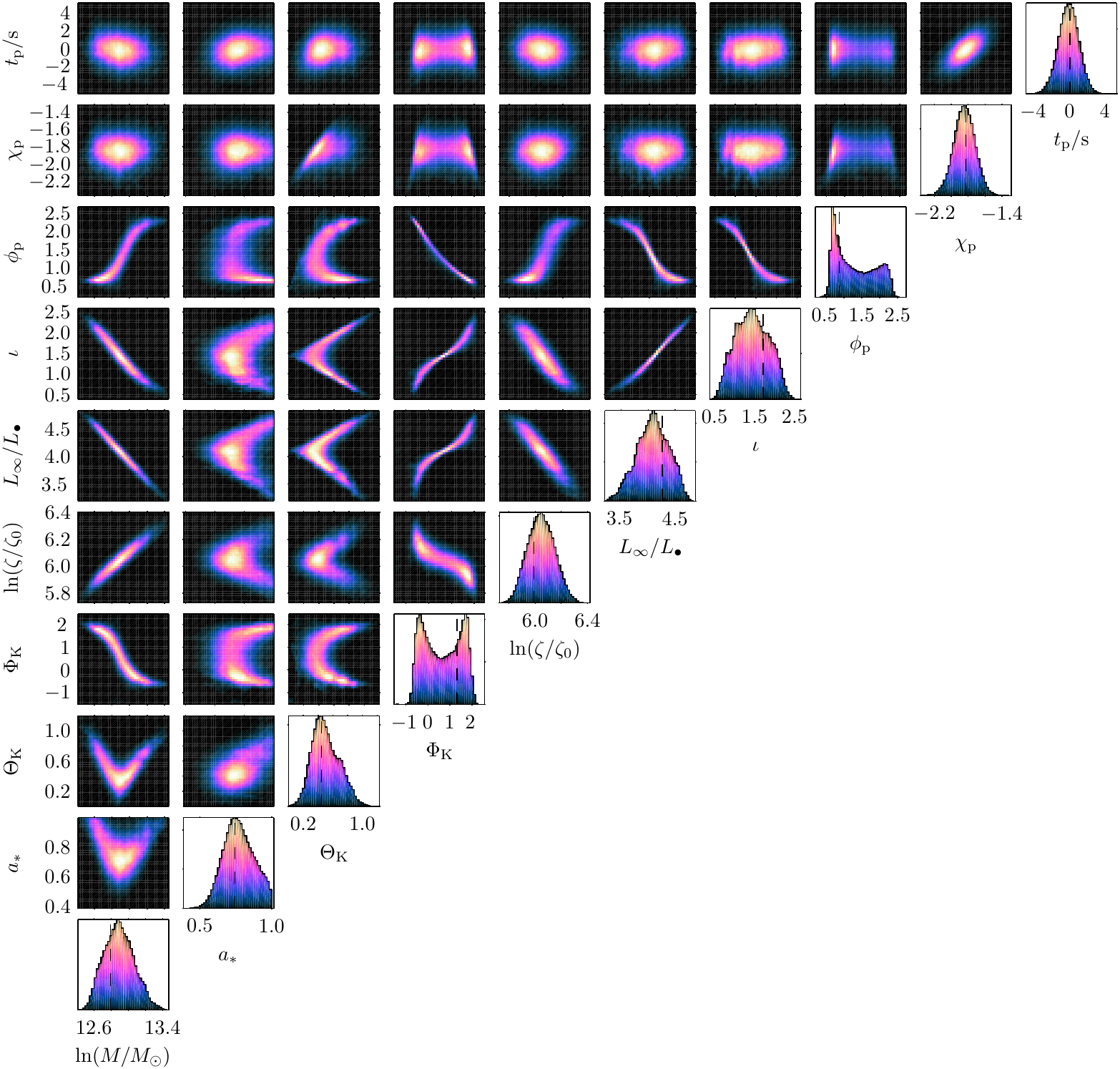}
\caption{Marginalised one- and two-dimensional posteriors. The conventions are the same as in \figref{MCMC-a}. These distributions begin to show the complicated shapes of degenerate distributions. The EMRB is from NGC 4395 and has $r\sub{p} \simeq 5.92 r\sub{g}$.\label{fig:MCMC-b}}
\end{center}
\end{figure*}

\begin{figure*}
\begin{center}
   \includegraphics[width=0.92\textwidth]{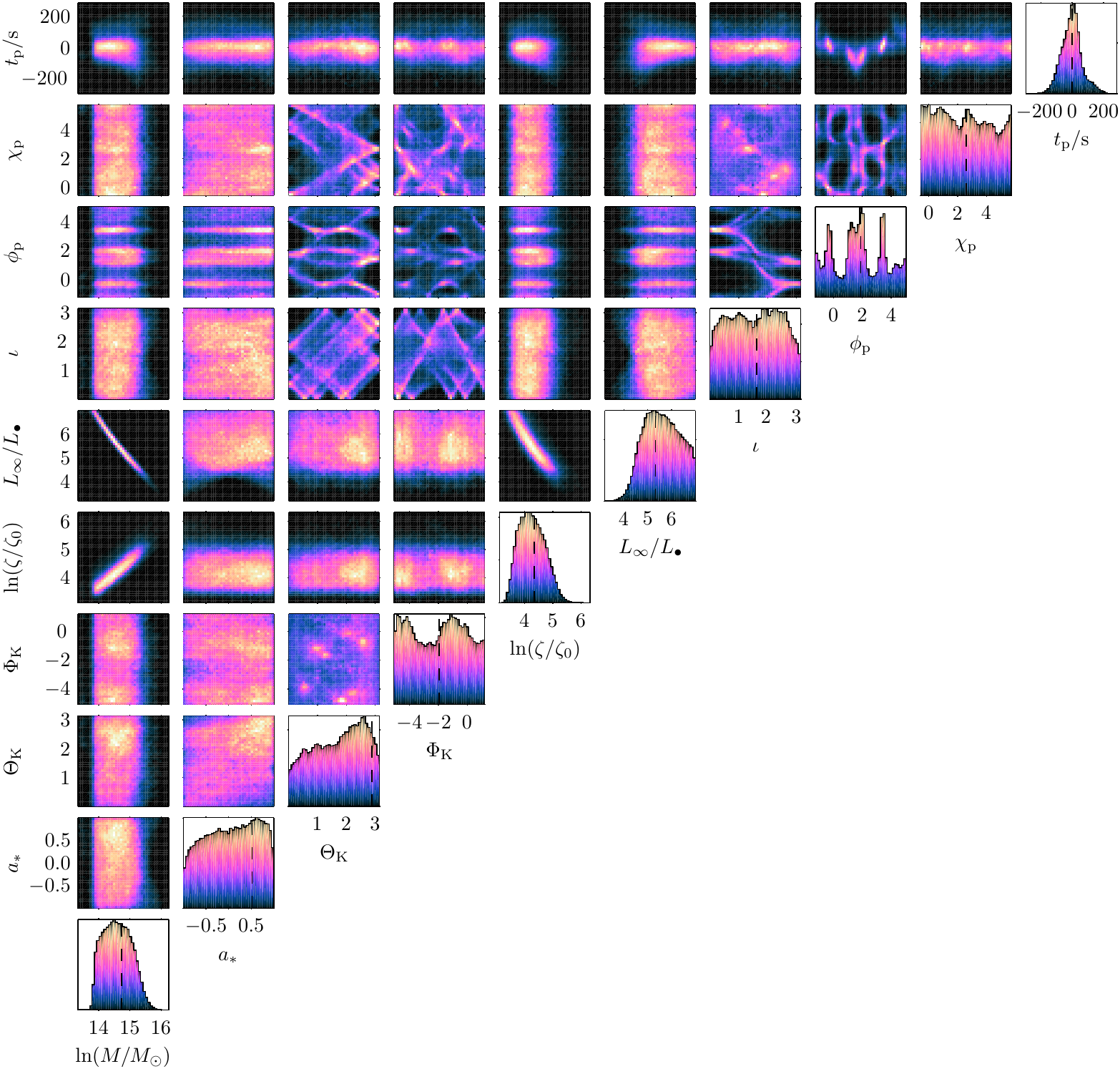}
\caption{Marginalised one- and two-dimensional posteriors. The conventions are the same as in \figref{MCMC-a}. These are the worst-case scenario distributions that are uninformative. The EMRB is from M32 and has $r\sub{p} \simeq 11.79 r\sub{g}$.}
\label{fig:MCMC-c}
\end{center}
\end{figure*}

The posteriors recovered from our MCMC show a wide variety of forms. There is a spectrum from well-formed Gaussians through elongated ellipsoids to complete covering of the parameter range. Some example results are shown in \figref{MCMC-a}, \ref{fig:MCMC-b} and \ref{fig:MCMC-c}.

\Figref{MCMC-a} shows the posterior for an EMRB from M32 with $r\sub{p} \simeq 5.53 r\sub{g}$. The distribution is well-defined and near Gaussian, although even in this best case the presence of degeneracies is clear. This example illustrates that it is possible to obtain good results, similar to those from the Galactic Centre, from extragalactic sources. Unfortunately, such tight distributions are not common in our sample.

\Figref{MCMC-b} shows the posterior for an EMRB from N4395 with $r\sub{p} \simeq 5.92 r\sub{g}$; it illustrates a more usual posterior. Typical posteriors are not Gaussian; the forms vary significantly, such that it is not possible to produce a standard shape. Non-Gaussianity manifests by the distributions broadening, developing curves and becoming banana-like. The degeneracies may evolve such that there are multiple modes.

\Figref{MCMC-c} shows the culmination of the deterioration of the posterior; it is for an EMRB from M32 with $r\sub{p} \simeq 11.79 r\sub{g}$. In this case, the distributions have extended to encompass the entire range for some parameters and so the EMRB is (near) useless. The posteriors show intricate degeneracies in some angular parameters. These are naturally periodic and demonstrate that near identical bursts can be produced through various rotations of the MBH and orbit. Such bursts are not informative and so are not of interest, but we include this example so that there is no illusion of all EMRBs having perfect posteriors.

\bsp

\label{lastpage}

\end{document}